\documentclass[]{aastex631}

\usepackage{hyperref}
\usepackage{natbib}
\usepackage{graphicx}
\usepackage{comment}
\usepackage{gensymb}
\usepackage{amsmath}
\usepackage{amssymb}

\begin{document}

\title{The Role of Spiral Arms in Galaxies II: Similarities Amid Diversity}

\author[0009-0000-3578-9134]{Bingqing Sun}
\affiliation{Department of Astronomy, University of Massachusetts Amherst, 710 North Pleasant Street, Amherst, MA 01003, USA}

\author[0000-0002-5189-8004]{Daniela Calzetti}
\affiliation{Department of Astronomy, University of Massachusetts Amherst, 710 North Pleasant Street, Amherst, MA 01003, USA}

\author[0000-0003-4569-2285]{Andrew J. Battisti}
\affiliation{International Centre for Radio Astronomy Research, University of Western Australia, 7 Fairway, Crawley, WA 6009, Australia}
\affiliation{Research School of Astronomy and Astrophysics, Australian National University, Cotter Road, Weston Creek, ACT 2611, Australia}
\affiliation{ARC Centre of Excellence for All Sky Astrophysics in 3 Dimensions (ASTRO 3D), Australia}

\begin{abstract}
The role of spiral arms in galaxies -- whether they enhance star formation efficiency or primarily act as material gatherers -- remains an open question. Observational studies have yielded ambiguous results, in part due to the choice of star formation rate (SFR) tracers and their inherent limitations.  These limitations are addressed here by applying multi-wavelength spectral energy distribution (SED) fitting to individual arm and interarm regions. We expand on our previous study of two galaxies to include six diverse galaxies, spanning over an order of magnitude in total stellar mass and factors of several in total SFR, for which spiral arms have been mapped. We find that the specific star formation rate (sSFR = SFR/M$_\textup{star}$) can be used as a proxy for the star formation efficiency (SFE=SFR/M$_\textup{gas}$), since the two quantities are directly proportional to each other in our regions. In our analysis of both tracers (sSFR and SFE) no significant difference is found the between arm and interarm regions, except for one galaxy (NGC 1097), supporting the gatherers scenario. 
\end{abstract}


\section{Introduction}\label{sec:intro}
Spiral arms are important features present in two-thirds of low-redshift galaxies \citep{Nair+2010}. However, the role spiral arms play in galaxy--wide  star formation remains uncertain \citep{Binney&Tremaine2008}.
Two basic scenarios have been proposed in the literature. In the ``trigger" scenario, spiral arms may enhance star formation efficiency by compressing gas and causing shocks, possibly reaching self-gravitating conditions. In the other scenario, the spiral arm acts as a ``gatherer'' that simply assembles the materials together. A review of the two scenarios can be found in \cite{DobbsBaba2014}.

Observations have so far yielded ambiguous results on whether one or the other scenario dominates in galaxy disks. Several studies, for instance, support the ``trigger'' scenario. \cite{LordYoung1990} reported a high contrast for the star formation rate, up to a factor of 2.3, between arm and interarm regions in the galaxy M51 in excess of the gas density contrast, indicating higher star formation efficiency in the arms. \cite{Cepa+1990} also support the trigger scenario, as they found a significant increase in the ionizing photon rate density relative to the underlying HI density within the arms of two grand-design spirals. \cite{Knapen+1996} defined the star formation efficiency parameter as the ratio of star formation rate to total (atomic+molecular) gas mass and found that this parameter is, on average, three times higher in spiral arms compared to the interarm regions in NGC 4321, concluding that star formation is triggered within the spiral arms. \cite{Seigar+2002} investigated both the trigger and gatherer scenarios and reported a difference of up to a factor of 2 in the H$\alpha$/K-band flux ratio in 20 spiral galaxies, concluding that star formation is likely triggered by shocks in spiral arms, under the assumption that shocks are a key driver of star formation.
\cite{Hitschfeld+2009} found smaller values of the Toomre Q-parameter in the spiral arms of M51, meaning that they are more likely to collapse under the gravitational pull; the authors conclude that this is expected as star formation occurs primarily in the arms. 
\cite{Rebolledo+2012} reported that some of the giant molecular clouds (GMCs) located in the arms of NGC 6946 have higher star formation efficiency than the rest of the GMCs. Finally, \cite{Yu+2021} analyzed a sample of 2226 nearby disk galaxies and found that both star formation rate (SFR) and specific star formation rate (${\rm sSFR} = {\rm SFR} / M_{\rm stars}$) increase with the strength of the spiral arms, providing further evidence in support of the trigger scenario.

Conversely, many other observational studies support the gatherer scenario, in which spiral arms primarily serve to accumulate gas material without directly triggering star formation. \cite{Elmegreen+1986} analyzed SFRs derived from H$\alpha$ and UV fluxes as a function of spiral arm type and Hubble type, finding no significant difference in the average SFRs, regardless of whether grand-design spiral structures are present or not.
\cite{Hart+2017} showed that the subsamples of galaxies with different spiral arm numbers show only marginal differences in their sSFR. \cite{Foyle+2010} reported that the interarm regions contribute at least 30\% of the SFR in spirals, but there is little enhancement in star formation efficiency (defined as the ratio of SFR to molecular gas mass, ${\rm SFE} = {\rm SFR} / M_{{\rm H}_2}$), less than 10\%, between arm and interarm regions in both grand-design and flocculent galaxies.
\cite{Kreckel+2016} found that the physical properties, including gas depletion time scales, of hundreds of arm and interarm HII regions that are traced by H$\alpha$ do not depend on location.  
The more recent work of \cite{Querejeta+2021} and \cite{Querejeta+2024} also support the gatherer scenario and conclude that the spiral arms do not boost the star formation efficiency. These extragalactic studies are complemented by studies of regions in the Milky Way that also find no significant enhancement in the SFEs of regions at different Galactic locations. 
\citet{Moore+2012} studied Milky Way massive young stellar objects and molecular clouds, reporting that source crowding within the arms can account for 60–80\% of the observed SFE (SFR/$M_{{\rm H}_2}$; where the SFR is estimated by the luminosity produced by IR-selected, massive young stellar objects and H II regions, and $M_{{\rm H}_2}$  is traced by $^{13}$ CO $J=1-0$) increase. 
\citet{Ragan+2018} showed that, while spiral arms tend to host the galaxy’s massive clusters, 92.4\% of the sample clumps in their catalog does not show enhanced star formation within the arms relative to other regions. Finally,  \cite{Urquhart+2021} found that the star formation processes in GMCs depend more on local environment rather than location within or outside of spiral arms. 

While observations have not yet yielded a final answer as to what role spiral arms play in star formation, most current simulations support the gatherer scenario. Simulations by \cite{Duarte-Cabral+2017} show that spiral arms cause giant molecular filaments to align and form clouds, but there is no increase in star-formation events after the gas enters the spiral arm. Similarly, \cite{Kim+2020} and \cite{Dobbs+2011} find that the presence of spiral arms enhances SFRs by factors of 2 or less, again supporting the gatherer scenario. 
Simulations of interacting galaxies show that SFRs do not increase when spiral arms are experiencing morphological changes due to the interactions \citep{Tress+2020}.

In an earlier work of this series \citep{Sun+2024}, we found an increase of only a factor of 1.1--2.1 in  SFE and $<$2 in sSFR in the arm regions relative to the interarm regions for two galaxies, NGC 628 and NGC 4321, also supporting the gatherer scenario. In that paper, we sought to overcome earlier limitations in deriving SFRs by using model fitting of multi--wavelength spectral energy distributions as opposed to the standard approach that uses measurements at one or two wavelengths only. As already pointed out in \citet{Sun+2024}, there are potential problems and complexities when these limited--wavelength SFR indicators are applied to regions within galaxies. 
\cite{Popescu+2005} finds that the FIR/UV ratios tend to be higher in interam regions relative to the arms, which can be interpreted as UV photons escaping from the spiral arms and being absorbed by dust, rather than enhanced star formation in the interarm regions. Studies of resolved stars support the conclusion that the interarm regions receive significant contribution to their UV emission from stellar streams out of arm regions \citep{Crocker+2015}.
The most widely used SFR indicator, the H$\alpha$ luminosity \citep{Kennicutt&Evans2012}, also has complications. Studies show that at least half of the diffuse, smooth H$\alpha$ in galaxies is due to leakage of ionizing photons from HII regions \citep{Thilker+2002, Oey+2007}, and there are large variations in the diffuse H$\alpha$ component from region to region. In some interarm regions, the diffuse, non--local component can represent up to 100\% of the total H$\alpha$ emission. Another common SFR indicator, the infrared emission from dust, also has its own complexities, as a significant portion of the dust emission, between 30\% and 80\%, is heated by older stellar populations which are not related to recent star formation \citep{Draine+2007, Calzetti+2007, Crocker+2013, Kirkpatrick+2014, Calapa+2014}.

The present work is a follow-up of our first paper \citep{Sun+2024}, in which we demonstrated our methodology by applying it to the data of two galaxies, NGC 628 and NGC 4321. We used the publicly available software Multi-wavelength Analysis of Galaxy Physical Properties \citep[MAGPHYS,][]{daCunha+2008} to perform fits of the UV--to--farIR SEDs of $\sim$1.5~kpc regions within those two galaxies, to derive SFRs and stellar masses. Previous work had shown that multi--wavelength SED fits derive internally consistent physical parameters, especially when applied to $\sim$kpc--sized regions, thereby mitigating issues associated with single- or dual-band recipes, especially in regions with very low sSFRs \citep{Hunt+2019, SmithHayward2018}. Here we extend our approach to a sample of four additional galaxies, for a total of six, that covers a larger range in the parameter space of stellar mass (factor $\sim$10) and SFR (factor $\sim$7) than our previous study. These are all the galaxies that satisfy our selection criteria, as described in Section \ref{sec:data}, and also have publicly available spiral arm masks, required for our analysis \citep{Querejeta+2021} . 
These galaxies cover a stellar mass range of $10^{9.61}$  M$_\odot$ to $10^{10.65}$ M$_\odot$, and cover a SFR range of 0.9 to 6.6 M$_\odot$yr$^{-1}$. 
By applying a consistent analysis across a diverse sample, we aim to identify common patterns—or similarities—amid the diversity of galaxy properties, and to assess whether spiral arms play a universal role in star formation.

In this work, we use both sSFR and SFE for our tests.
\cite{Seigar+2002} argue that if density waves primarily act as gatherers, they should concentrate all types of disk material including stars, to the same extent. However, if spiral density waves significantly trigger star formation, the arm-to-interarm contrast in the star formation rate 
would be substantially higher than that in the old stellar population. Similarly, \cite{Hart+2017} argue that the sSFR in star-forming galaxies  clearly traces how efficiently gas is converted to stars.
\cite{Saintonge+2011} showed that the sSFR of galaxies tightly correlates with their molecular gas depletion timescale (inverse of the SFE), providing a strong basis for our use of the  sSFR as a tracer of the star formation enhancement in spiral arms. While the work by \cite{Saintonge+2011} present results for  whole galaxies, we demonstrate in Section~\ref{sec:results} that the sSFR--SFE correlation holds at the $\sim$kpc region level, thus enabling us to draw conclusions based on the trends of both sSFR and SFE.

The structure of this paper as follows. In section \ref{sec:data} and \ref{sec:method} we describe the imaging data used in this study and the processing methods applied to the datasets. We present and discuss the results for the six galaxies in Section \ref{sec:results}. Section \ref{sec:summary} provides a brief summary of this paper.

\section{Data}\label{sec:data}
The sample consists, in addition to NGC\,628 and NGC\,4321 studied in \cite{Sun+2024}, of four additional galaxies whose characteristics are listed in Table~\ref{table:aperture}. We require all galaxies to have imaging data from the far--UV (FUV) to the far--IR (FIR), to be able to fit not only the UV--optical--nearIR stellar emission, but also the mid/far infrared dust emission. Fitting the dust emission is key to recover the dust---attenuated stellar emission and derive unbiased SFRs and stellar masses, under the assumption of energy balance between absorption and emission. The minimum imaging suite we require for each galaxy is to have archival data from the following facilities/surveys: the Galaxy Evolution Explorer (GALEX; 2 bands) in the UV; optical imaging (4+ bands, typically SDSS); the Two Micron All-Sky Survey (2MASS; 3 bands) in the near--IR; the Spitzer Infrared Array Camera (IRAC; 4 bands) and the Multiband Imaging Photometer on Spitzer (MIPS; 3 bands), and the Herschel Photodetector Array Camera and Spectrometer (PACS; 2 or 3 bands) and the Spectral Photometric Imaging Receiver (SPIRE 250 $\mu$m; 1 band) to cover the range 3--250~$\mu$m. We also require the images to cover the entire extent of each galaxy and to have inclination $i \leq 60 \degree$, to reduce line-of-sight confusion. Here NGC\,4254 and NGC\,4579 have 5 SDSS bands in the optical, while NGC\,1097 and NGC\,1512 have 4 optical bands (B, V, R and I) from other sources. All our galaxies have a minimum of $19$ bands each, covering the wavelength range 0.15--250~$\mu$m. All archival images are available from the NASA Extragalactic Database (NED\footnote{http://ned.ipac.caltech.edu/}).

\begin{table}[tbp]
\begin{center}
\begin{tabular}{ ccccccccccc } 
\hline
Name & RA   &  Dec  & 2a & 2b  & PA  & Distance & Inclination & Morphology & Stellar Mass & SFR\\
&(deg) &(deg) &($^{\prime\prime}$)&($^{\prime\prime}$) &($\degree$)&(Mpc) &  ($\degree$) & &  (M$_\odot$) & (M$_\odot$yr$^{-1}$)\\
\hline
NGC 628 & 24.173946 & 15.783662 & 586.0 & 538.7 & 90 & 9.77& 5 & SA(s)c & 9.68 & 0.15\\
NGC 4321 & 185.728463 & 15.821818 & 558.0 & 483.0 & 40 & 15.20 & 27 & SAB(s)bc & 10.36 & 0.58\\
NGC 1097 & 41.579375 & -30.274889 & 380.0 & 585.0 & 122 & 14.20 & 48.6 & SB(s)b& 10.65 &0.82\\
NGC 1512 & 60.976167 & -43.348861 & 754.0 & 560.0 & 83 & 11.70 & 42.5 & SB(r)a & 10.10 & -0.04\\
NGC 4254 & 184.706682 & 14.416509 & 519.0 & 420.0 & 60 & 13.90 & 29 & SA(s)c & 9.61 & 0.63\\
NGC 4579 & 189.431342 & 11.818194 & 270.0 & 256.0 & 90 & 16.70 & 39 & SAB(rs)b & 9.96 & 0.01\\
\hline
\end{tabular}
\end{center}
{References for the distances: \citet{Tully+2013} for NGC\,4254 and NGC\,4579, \citet{Sabbi+2018} for NGC\,1512, and \citet{Nasonova+2011} for NGC\,1097. References for inclinations: \citet{Menendez-Delmestre+2007} for NGC\,1512, NGC\,4254, NGC\,4579, and \citet{Lang+2020} for NGC\,1097. References for the analogous parameters of NGC\,628 and NGC\,4321 are in \citet{Sun+2024}.}
\caption{List of sample galaxies and their parameters. Right Ascension(RA) and Declination(Dec) of the galactic center in degrees, galaxy morphology, distances, and inclination angles are obtained from NED. The photometric apertures (ellipses with sizes 2$\times$semi-major axis (2a) and 2$\times$semi-minor axis (2b)) and position angles (PA) adopted for the global photometry are also listed \citep[see Section \ref{SED_fitting} for details]{Dale+2017}.  Stellar masses and SFRs are in log scale.}
\label{table:aperture}
\end{table}

As our approach consists of dividing each galaxy into spaxels $\sim 1.5$~kpc in size, we require all galaxies to be 
within distance $\leq 20$ Mpc.
Each spaxel is then assigned to either an arm or an interarm region (see below), and its multi--wavelength SED fit with MAGPHYS to derive the SFR and stellar mass. As assessed in \citet{Sun+2024}, a 1.5~kpc scale resolution for each galaxy is sufficient to easily separate arm from interarm regions, and obtain enough spaxels in each region for statistical analyses.  

\section{Method}\label{sec:method}

\subsection{Image Processing and Spiral Arm identification}
We adopt the same image processing and identification of spiral arm/interarm spaxels approach of our previous work \citep{Sun+2024}, which we briefly summarize here. We apply the methodology to NGC 1097, NGC 1512, NGC 4254, and NGC 4579. The relevant physical quantities for the remaining two galaxies, NGC 628 and NGC 4321, are from \citet{Sun+2024}.

First, we mask the bright foreground stars and background galaxies that fall within the body of each galaxy to avoid contamination. We use the masks produced by D. Dale (private communication) using both visual inspection and a multiwavelength analysis for all galaxies in our sample except NGC\,1097. For this galaxy, we produce our own stellar masks from the B and I band images, using a similar approach as D. Dale's. 

We subsequently align the images of each galaxy to the reference frame of the Spitzer/IRAC 3.6 $\mu$m band, resampling all images to the IRAC pixel scale of 0.$^{\prime\prime}$75 using the software SWarp. 
Background subtraction is performed during resampling using a grid size much larger than the default in SWarp, to avoid over-subtraction of galaxy fluxes.
Next, the resampled images are convolved to a common point-spread function (PSF), using the PSF of the lowest-resolution image in our analysis, the SPIRE 250 image(FWHM $\sim$ 18$^{\prime\prime}$), using the convolution kernels from \cite{Aniano2011}.
Finally, all images are resampled to spaxels with sizes of 18$^{\prime\prime}$, which is consistent with the size of the SPIRE 250 PSF, so that each spaxel corresponds to a region of $\sim$ 1.5 kpc.

Following \cite{Sun+2024}, we separate the arm and interarm regions by applying the masks from \cite{Querejeta+2021}. These authors define five different galaxy environments including spiral arms, centers, bars, interarm regions and disks. Spaxels with more than half of their area in the arm mask are attributed to the arm region. \cite{Sun+2024} showed that this choice is robust against variations in the covering fraction as long as the fraction is at least 50\%  in the arm region. We also exclude the central regions and bars, as specified by the mask, as we did for NGC 628 and NGC 4321. All remaining spaxels within the disk are classified as interarm regions. 
For NGC 1097 and NGC 1512, the `disk' defined by the \cite{Querejeta+2021}'s masks covers a smaller FOV than the global aperture from \cite{Dale+2017} does, so we adopt the size of the mask's `disk' as our aperture for the global photometry of these two galaxies.

\subsection{SED Fitting}\label{SED_fitting}
Before constructing and fitting the SEDs of individual spaxels, we conduct a consistency check by comparing the global fluxes of all six galaxies with the global photometric results from \cite{Dale+2017}. 
We refer to their apertures for the photometry of NGC 4254. We adjusted the aperture size of NGC 4579 to account for the limited coverage of our images, which is slightly smaller than the area of the galaxy (from $325^{\prime\prime}\times 271^{\prime\prime}$ to $270^{\prime\prime}\times 256^{\prime\prime}$, with the same PA).
For NGC 1512 and NGC 1097, we reduced the aperture sizes in order to match the area of the galaxy disks as defined by the mask that we use to identify spiral arm regions. These apertures are used for both global photometry and the selection of spaxels for fitting and analysis.
Our global aperture sizes and position angles are listed in Table \ref{table:aperture}. We determine that our flux measurements are in good agreement with those in \cite{Dale+2017}, within $\leq 10$ \% difference. 

Each spaxel SED consists of flux measurements in at least 19 bands (see earlier sections) plus uncertainties. The uncertainties are calculated following the method described in \cite{Sun+2024}. Basically, we plot the distribution of the spaxel values for the entire image at each wavelength, and then fit a single Gaussian to the left side of the peak. The 1-$\sigma$ width of this Gaussian is then adopted as the flux uncertainty for each spaxel.

The steps we follow for the SED fitting are the same as our previous work, in which we used the fitting code MAGPHYS \citep{daCunha+2008}. MAGPHYS includes stellar population models with a range of metallicities, and allows multiple bursts of star formation to be superimposed on an exponentially declining SFR, enabling some flexibility in the recovered star formation history. Energy conservation between dust absorption in the UV--optical and dust emission in the IR is also a basic assumption of MAGPHYS, which \citet{SmithHayward2018} have shown to hold down to $\sim$1~kpc scale in galaxies. 
Same as in \citet{Sun+2024}, we select all spaxels with a signal-to-noise ratio (S/N) greater than or equal to 3 in the SPIRE 250 image for both  NGC 4254 and NGC 4579. This threshold ensures that only regions with reliable flux measurements are included in the analysis. In the case of NGC 1097 and NGC 1512, we need to relax this assumption as the SPIRE 250 images of these galaxies generally have lower S/N values; for these two galaxies, we perform SED fits on all spaxels with S/N $\geq$ 3 in IRAC 3.6 $\mu$m images; with this assumption, all other bands, except SPIRE 250, have higher S/N. We have also tested this approach on the other two galaxies, NGC 4254 and NGC 4579. When selecting spaxels with S/N $\geq$ 3 in the IRAC 3.6 $\mu$m images, we find that the conclusions remain unchanged. Therefore, to ensure that all bands are well detected, we retain the S/N $\geq$ 3 threshold in the SPIRE 250 $\mu$m band for these two galaxies.

\subsection{Gas Maps}
Similarly to what was done in \citet{Sun+2024} and other works, we estimate the molecular gas mass (M$_\textup{gas}$) and star formation efficiency (SFE) for NGC 1097, NGC 1512, NGC 4254 and NGC 4579, using the CO $J = 2\rightarrow 1$ maps from PHANGS-ALMA survey \citep{Leroy2021a}. We choose the strict mask mom0 images among the PHANGS data release \footnote{https://www.canfar.net/storage/vault/list/phangs/RELEASES/PHANGS-ALMA/} because they only include signal with high confidence in the data cubes. These maps generally recover less total flux than less strict criteria, but this is not an issue for our analysis, because we focus on relative comparisons between arm and interarm regions. We adopt a 
$J = 2 \rightarrow 1$ to CO $J = 1 \rightarrow 0$ conversion factor R$_{21}$ = 0.65,
and CO-to-H$_2$ conversion factor $\alpha_\textup{CO,MW}$ = 4.3 M$_\odot$/pc$^2$/(K km/s)\citep{Leroy2021a}. These images have pixel size of  0.5$^{\prime\prime}$. We follow the same processing steps listed in Section 3.1 to match the resolution of data.

To estimate the uncertainties in the M$_\textup{gas}$ and SFE of each spaxel, we use the corresponding emom0 maps from the PHANGS data release. 
For each galaxy, the emom0 maps are resampled to the same IRAC 3.6 frame as the other bands, while keeping the pixel scale unchanged. 
The final uncertainties of M$_\textup{gas}$ and SFE for each 18$^{\prime\prime}$ spaxel are then obtained by propagating the errors of all small spaxels that fall within its area.

\section{Results and Discussion}\label{sec:results}

\subsection{Statistics on Individual Galaxies}

\begin{figure}
    \centering
    \includegraphics[width=1\linewidth]{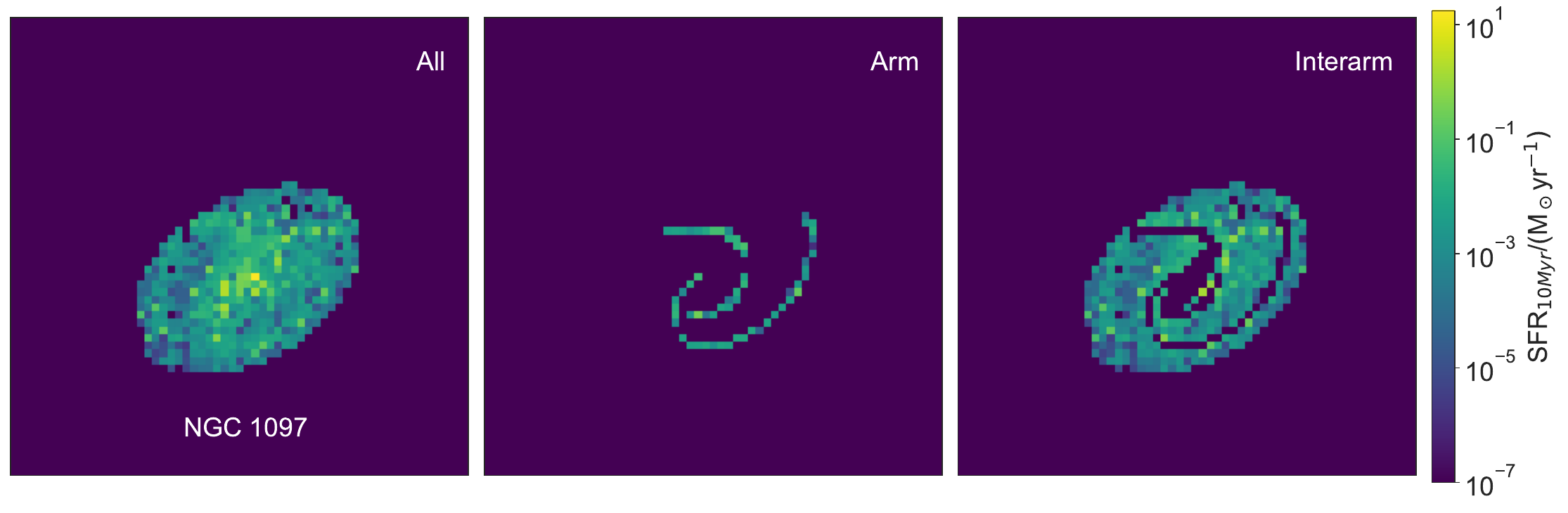}
    \includegraphics[width=1\linewidth]{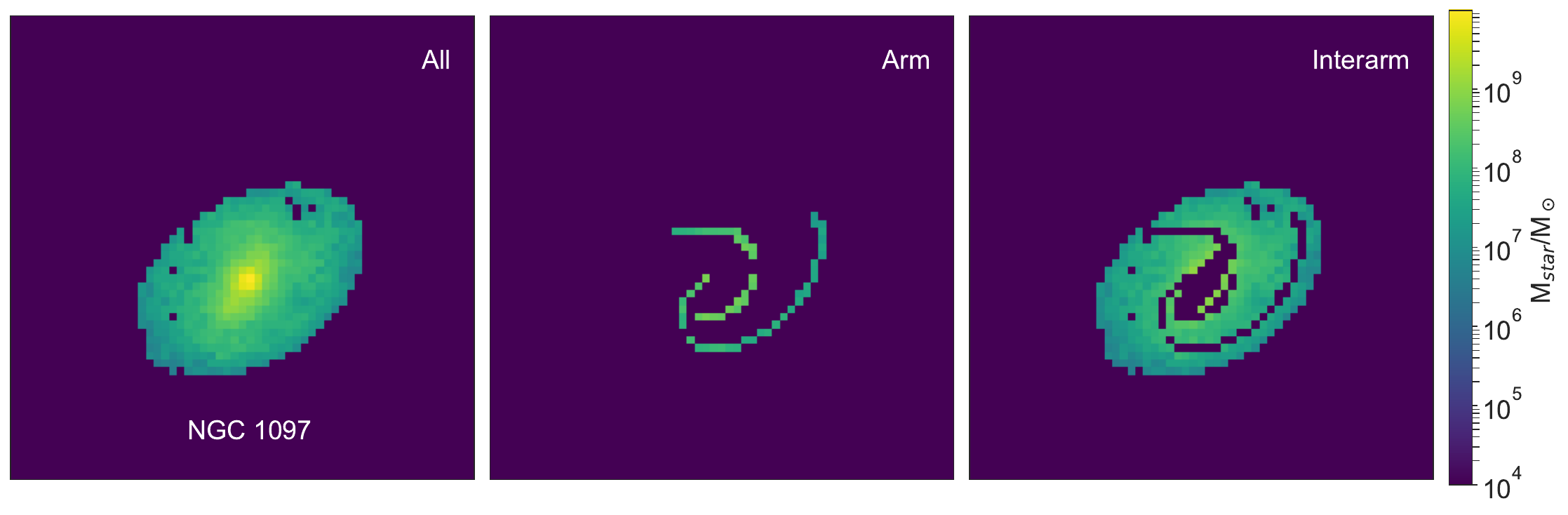}
     \includegraphics[width=1\linewidth]{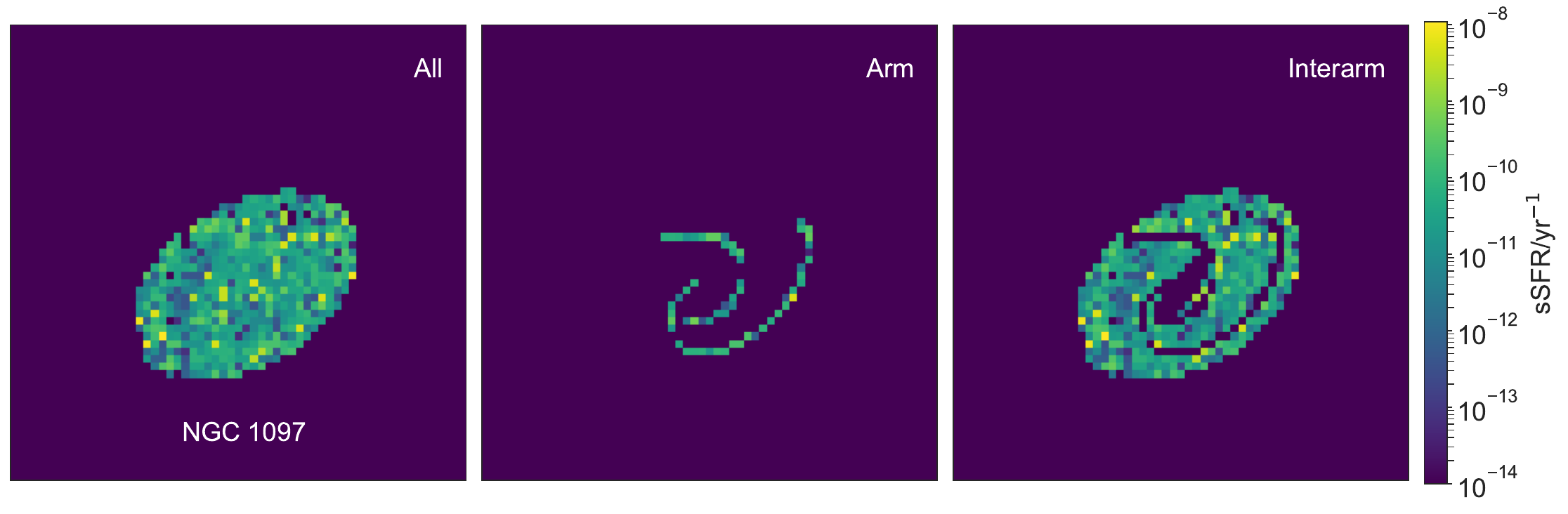}
    \caption{Parameter maps for NGC 1097. From upper to bottom rows: SFR; stellar mass; and specific star-formation rates (sSFR). In each row, the three panels from left to right are: best-fit values of all spaxels with $S/N \geq 3$ in IRAC 3.6 $\mu$m band (including include bulge spaxels); arm spaxels; and interarm spaxels, with the bulge spaxels excluded.}
    \label{fig:ngc1097-param-maps}
\end{figure}

In Figure \ref{fig:ngc1097-param-maps}
we show the maps of the best-fit physical parameters, SFR, M$_{star}$ and sSFR, derived for NGC 1097 from the  SED fits of all spaxels that are above the S/N threshold discussed earlier. We clearly observe the expected decrease of M$_\textup{star}$ per spaxel from the center to the outskirts of the galaxy. The parameter maps for the remaining three galaxies, NGC 1512, NGC 4254 and NGC 4579, are shown in Appendix~\ref{appendixA}. 
For NGC 628 and NGC 4321, the parameter maps can be found in \cite{Sun+2024}. Two of the shown parameters, M$_\textup{star}$ and SFR, are direct outputs of MAGPHYS. 

In order to quantify the differences between the distributions of physical parameters of the arm and interarm regions, we plot the distributions of parameter values, calculate the medians, and perform the two-sample Kolmogorov-Smirnov (K-S) test for each galaxy and for each parameter analyzed here: SFR, M$_\textup{star}$, sSFR, M$_\textup{gas}$, and SFE. The summary of the median values and the corresponding median ratios between arm and interarm regions is provided in Table \ref{table:individual_galaxies}, where we also include the p-values from the K-S test. The distributions of sSFR, M$_\textup{star}$ and SFR for each galaxy are shown in Appendix~\ref{appendixA}. 

For all galaxies, we find larger median values of the SFR (averaged over the most recent 10~Myr, which is shorter than the typical crossing time of spiral arms) and of M$_\textup{star}$ in the  spiral arm spaxels than in the interarms, as expected, since spiral arms tend to have higher levels of star formation per unit area than interarm regions.
For the sSFR, none of the galaxies show significant differences between the medians of the arm and the interarm regions, with the median ratio being $\lesssim 2$. 
Furthermore, the sSFR distributions for the arm and interarm regions in all galaxies have p-values $\geq 0.05$ in the two-sample K-S test, indicating no statistically significant differences; the exception is NGC 1097, which has a p-value of 0.0024, corresponding to a difference of 3.04~$\sigma$. 
The median arm/interarm ratio for  M$_\textup{gas}$ is between 1.2 to 2.2 for four  of the six galaxies; conversely NGC 4579 shows a ratio of medians $<$1, which indicates a higher gas mass per spaxel in the interarm regions than in the arms. We do not report the statistics of NGC 1512, because there are too few well-detected spaxels in the gas maps. For what concerns the SFE, all ratios of medians are between $\sim$0.6 and $\sim$2.1, but, with the exception of NGC628, none of the differences between arm and interarm regions are statistically significant.

\begin{table}[tbp]
\centering
\begin{tabular}{ l|c|c|c|c|c|c } 
\hline
 & NGC 1097 & NGC 1512 & NGC 4254 &  NGC 4579 & NGC 628 & NGC 4321 \\
\hline
\# Arm Spaxels & 60 & 33 & 43 & 25 & 147 & 66\\
\# Interarm Spaxels & 437 & 149 & 184 & 112 & 381 & 269\\
\hline
$M_\textup{star}$, Arm & 7.984 & 7.802 & 8.681 & 8.811 & 7.384 & 8.276\\
$M_\textup{star}$, Interarm & 7.673 & 7.268 & 8.141 & 8.520 & 6.994 & 8.053\\
Median Ratio & 2.044 & 3.419 & 3.468 & 1.957 & 2.452 & 1.671\\
p-value & 1.260e-4 & 2.674e-6 & 5.200e-10 & 7.445e-8 & 1.032e-8 & 6.175e-3\\
\hline
SFR$_\textup{10Myr}$, Arm & -2.263 & -2.799 & -1.215 & -2.036 & -2.668 & -1.957\\
SFR$_\textup{10Myr}$, Interarm & -2.829 & -3.359 & -1.967 & -2.453 & -3.209 & -2.370\\
Median Ratio & 3.691 & 3.628 & 5.652 & 2.612 & 3.385 & 2.587\\
p-value & 9.268e-7 & 0.210 & 5.675e-5 & 0.028 & 9.033e-10 & 1.588e-3\\
\hline
sSFR$_\textup{10Myr}$, Arm & -10.209 & -10.609 & -10.054 & -10.859 & -10.099 & -10.264\\
sSFR$_\textup{10Myr}$, Interarm & -10.516 & -10.789 & -10.089 & -11.154 & -10.265 & -10.497\\
Median Ratio & 2.030 & 1.513 & 1.084 & 1.971 & 1.468 & 1.709\\
p-value & 0.0024 & 0.150 & 0.800 & 0.114 & 0.064 & 0.050\\
\hline
\hline
\# Arm Spaxels (CO) & 22 & 3$^{(a)}$ & 31 & 18 & 145 & 64\\
\# Interarm Spaxels (CO) & 24 & 3 $^{(a)}$ & 66 & 31 & 361 & 259\\
\hline
$M_\textup{gas}$, Arm & 7.179 & ...  & 7.588 & 7.121  & 6.516 & 7.135\\
$M_\textup{gas}$, Interarm & 7.069 & ... & 7.244 & 7.325  & 6.375 & 6.894 \\
Median Ratio & 1.285 & ... & 2.209 & 0.625 & 1.383 & 1.740\\
p-value & 0.136 & ... & 0.024 & 0.0756 & 0.001 & 0.048\\
\hline
SFE, Arm & -9.116 & ... & -8.589 & -9.060 & -9.319& -8.752\\
SFE, Interarm & -8.877& ... & -8.716  & -9.201 & -9.643 & -8.799\\
Median Ratio & 0.577 & ... & 1.340 & 1.383 & 2.108 & 1.113\\
p-value & 0.104 & ... & 0.479 & 0.793 & 7.204e-5 & 0.368\\
\hline 
Spaxel size (kpc$^2$)  & 1.535 & 1.042 & 1.471& 2.124 & 0.727& 1.759 \\
Deprojected size (kpc$^2$)  & 3.511 & 1.918 & 1.923 & 3.516 & 0.732 & 2.216\\
\hline
\end{tabular}
\caption{The number of spaxels, the medians and median ratios of the best fit parameters for each galaxy, divided between arm and interarm spaxels. The medians are in log scale, while the ratios are in linear scale. Units for the physical parameters (all in log$_{10}$ scale) are: M$_{\odot}$ for M$_\textup{star}$ and M$_\textup{gas}$, M$_{\odot}$/yr for SFR, and yr$^{-1}$ for sSFR and SFE.\\
Note $^{(a)}$: We do not report the statistics of NGC 1512 because there are too few surviving spaxels.}
\label{table:individual_galaxies}
\end{table}

\begin{figure}
    \centering
    \includegraphics[width=1\linewidth]{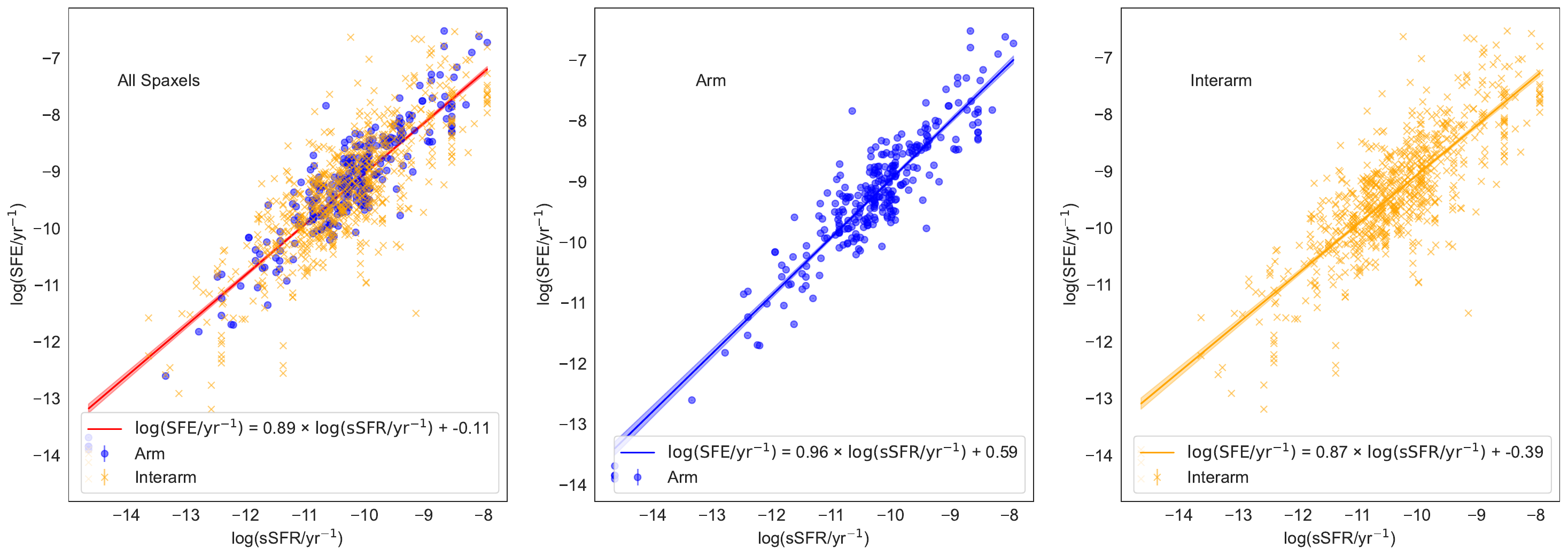}
    \caption{SFE vs sSFR in all galaxies. Left: all spaxels in six galaxies. Middle: all arm spaxels in six galaxies. Right: all interarm spaxels in six galaxies. The straight lines are the linear fit of log(sSFR/yr$^{-1}$) and log(SFE/yr$^{-1}$). }
    \label{fig:SFE_vs_sSFR_combined}
\end{figure}

So far, we have presented results separately for sSFR and SFE, but \citet{Saintonge+2011} showed that these two parameters are correlated in local galaxies. We now test whether a correlation between these two quantities also exists at the level of $\sim$kpc--sized regions. Figure \ref{fig:SFE_vs_sSFR_combined} shows the SFE versus sSFR of all valid spaxels across the six galaxies. Relations for individual galaxies are reported in Appendix~\ref{appendixC}. 
We find tight linear relations between log(sSFR/yr$^{-1}$) and log(SFE/yr$^{-1}$), in all arm spaxels, all interarm spaxels, and all spaxels combined together. The best-fit slopes and intercepts, derived using the hierarchical Bayesian algorithm LINMIX \citep{Kelly2007}, are shown in each panel of the plot. We conclude that results derived from sSFR are interchangeable with those derived from SFE, and can be directly compared with previous results that use either as tracers.

\subsection{Combining Results across Six Galaxies}
We now analyze the distributions of all physical parameters (M$_\textup{star}$, M$_\textup{gas}$, SFR, sSFR, and SFE) across a `super-galaxy' obtained by combining the spaxels from all six galaxies. To combine the data from the galaxies into a unified analysis, we normalize each parameter within each galaxy by dividing it by the average value across the disk (including both arm and interarm spaxels, but excluding the bulge) of that galaxy. For the specific star formation rate (sSFR), the average disk sSFR is computed as the ratio of the total star formation rate (SFR) of the disk to the stellar mass of the disk, [SFR(disk)/M$_\textup{star}$(disk)].

All of our galaxies have a spaxel size of 18$^{\prime\prime}$. However, they have non-negligible, and different, inclination angles. To ensure a consistent comparison of galaxy areas when combining the data, we correct for this inclination effect. 
For each parameter 
in each galaxy, we first normalize it by dividing by the average value of that parameter across the entire galaxy. We then multiply the result by cos $i$, where $i$ is the galaxy’s inclination angle.

\begin{figure}
    \centering
    \includegraphics[width=0.6\linewidth]{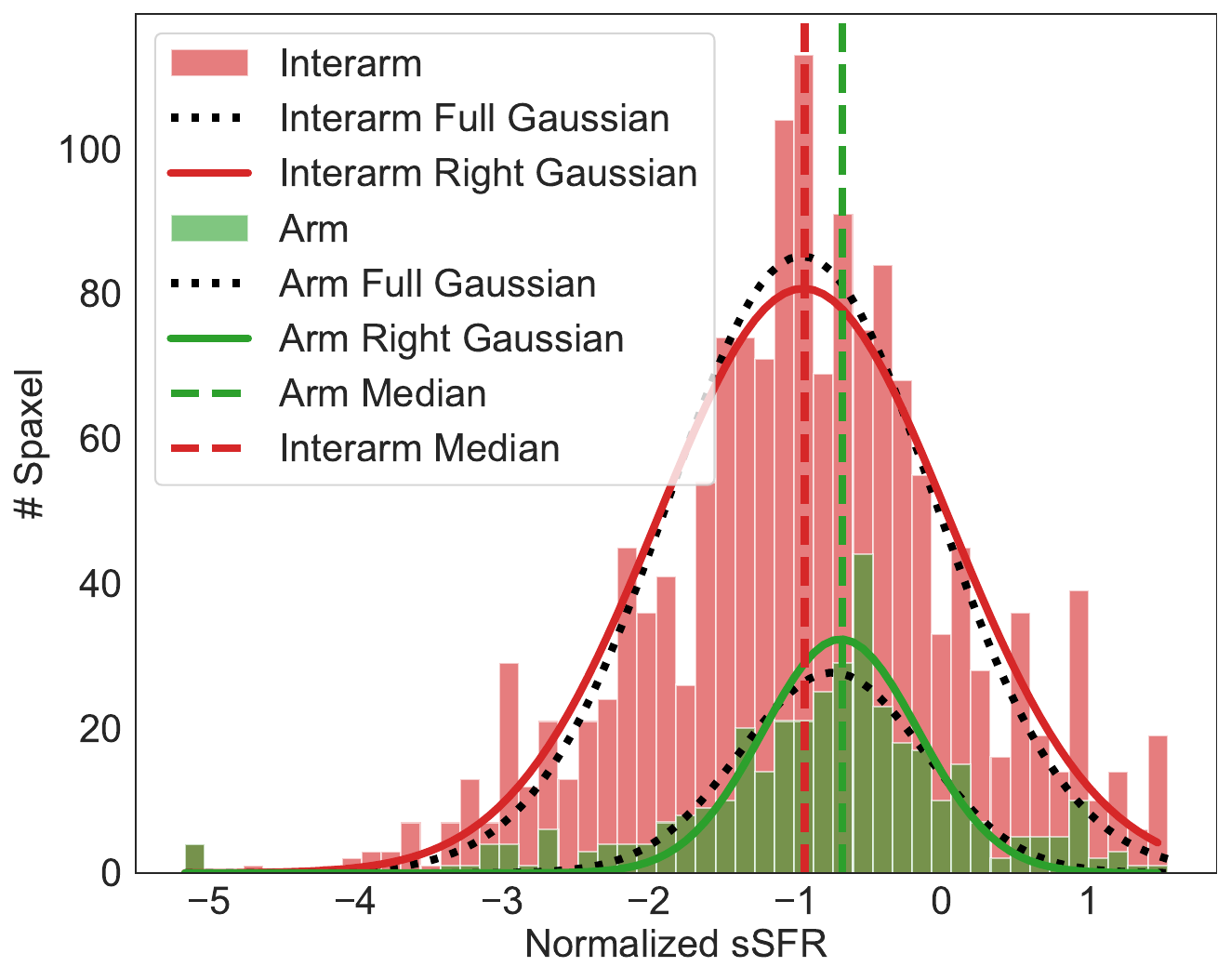}
    \caption{sSFR distributions for arm and interarm spaxels across six galaxies.
    The median values are -0.681 (arm) and -0.937 (interarm). The median ratio is  1.800. The black dotted lines are the gaussian fit of all arm/interarm spaxels. The solid lines are the gaussian fit of the right side of the distribution, centered on the median values, using the same color coding of the two distributions.}
    \label{fig:sSFR_six_galaxies}
\end{figure}

\begin{figure}
    \centering
    \includegraphics[width=0.47\linewidth]{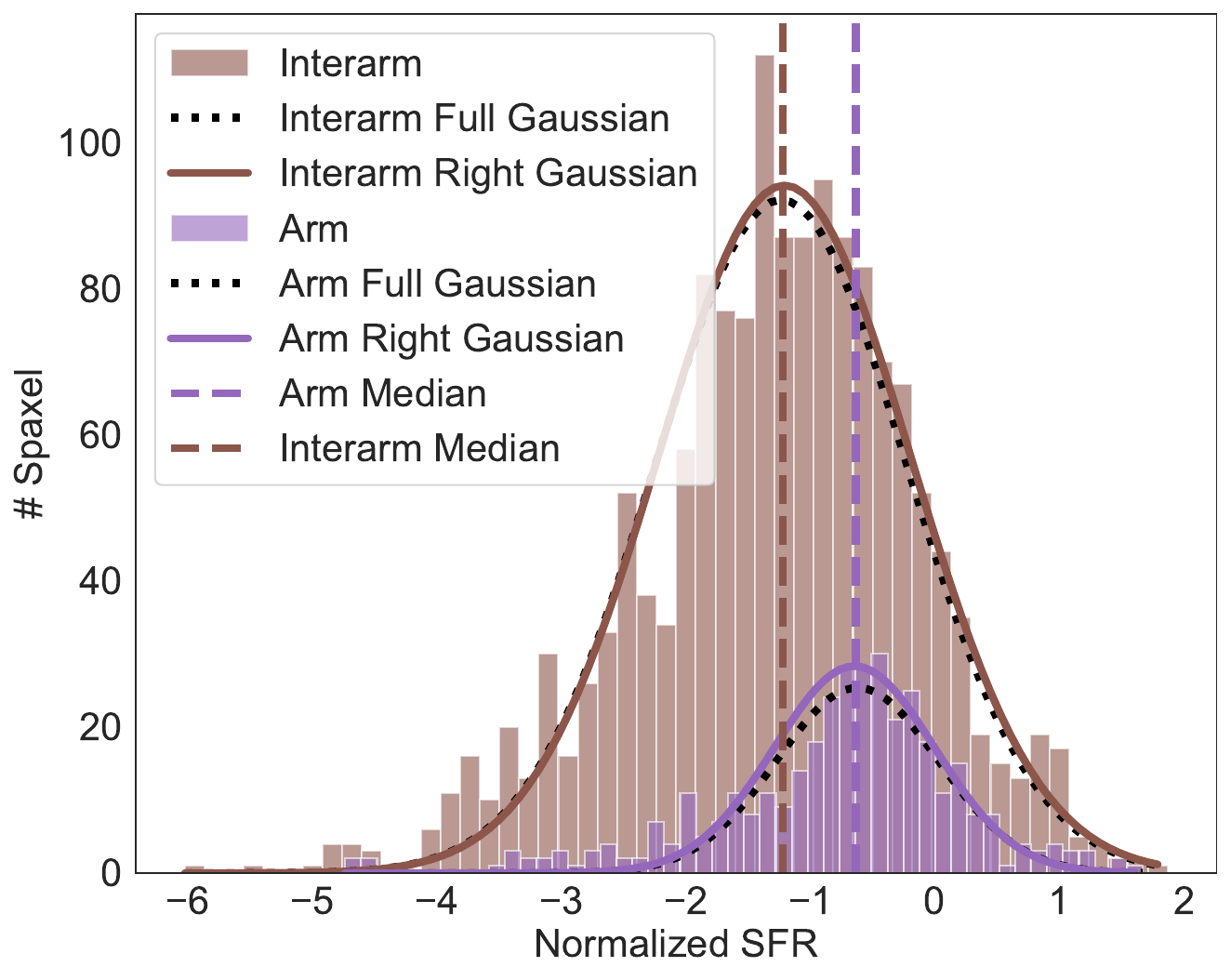}
    \includegraphics[width=0.47\linewidth]{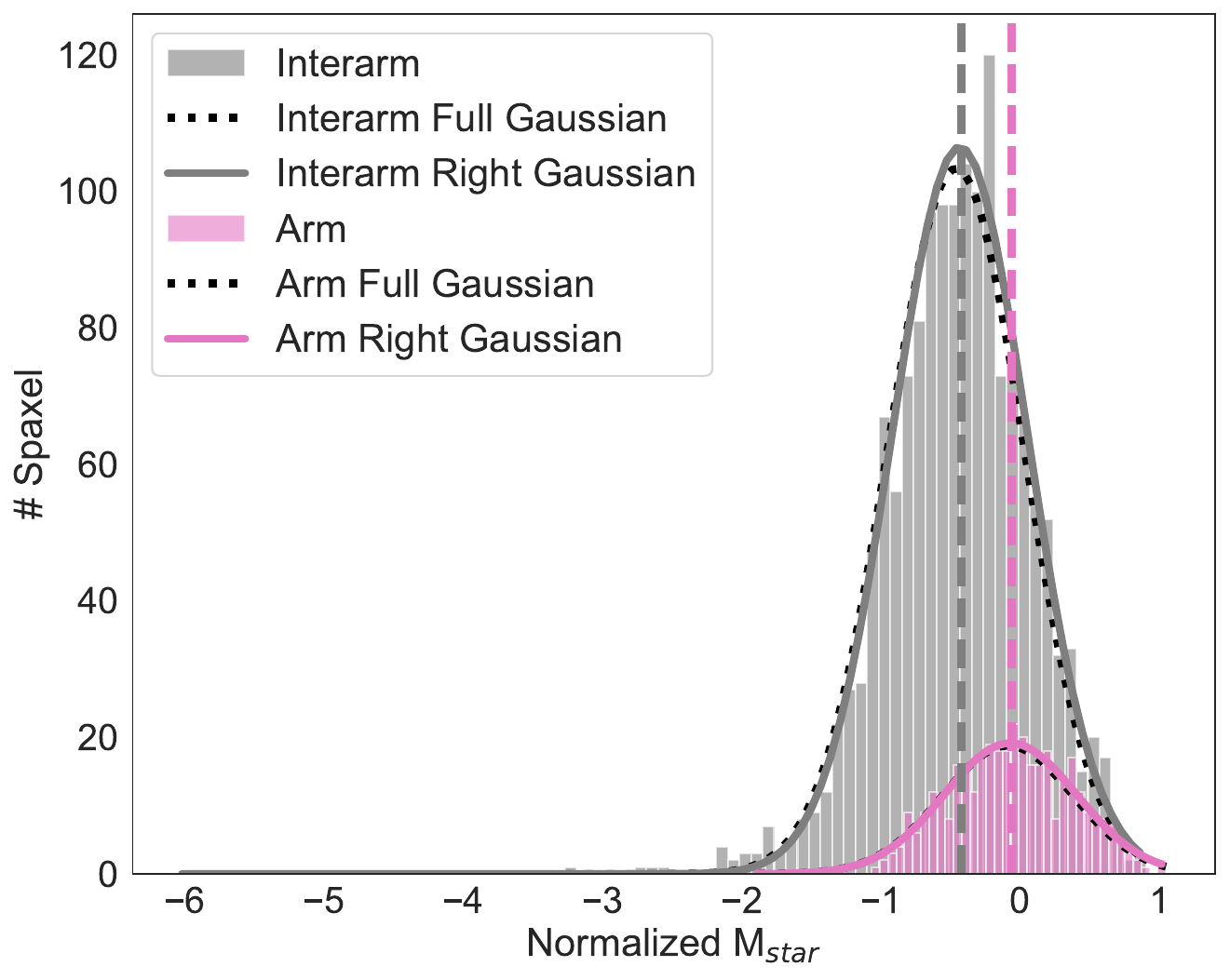}
    \caption{SFR and M$_\textup{star}$ distributions for arm and interarm spaxels across six galaxies.
    Left: Normalized SFR, with median log values of -0.636 (arm) and -1.218 (interarm), yielding a median ratio of 3.823.
    Right: Normalized M$_\textup{star}$, with median log values of -0.061 (arm) and -0.424 (interarm), corresponding to a median ratio of 2.304.
}
    \label{fig:SFR_Mstar_six_galaxies}
\end{figure}

\begin{figure}
    \centering
    \includegraphics[width=0.47\linewidth]{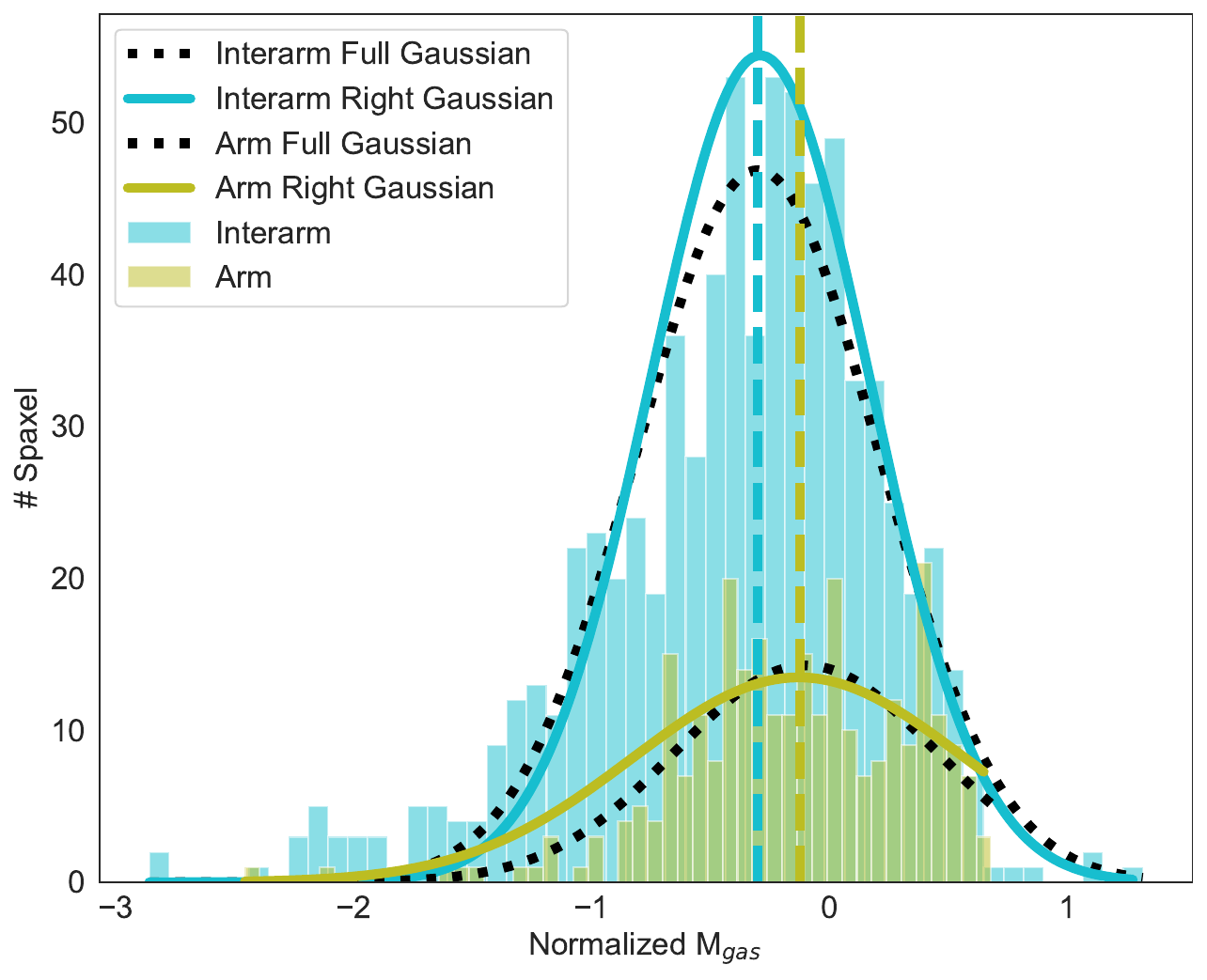}
    \includegraphics[width=0.47\linewidth]{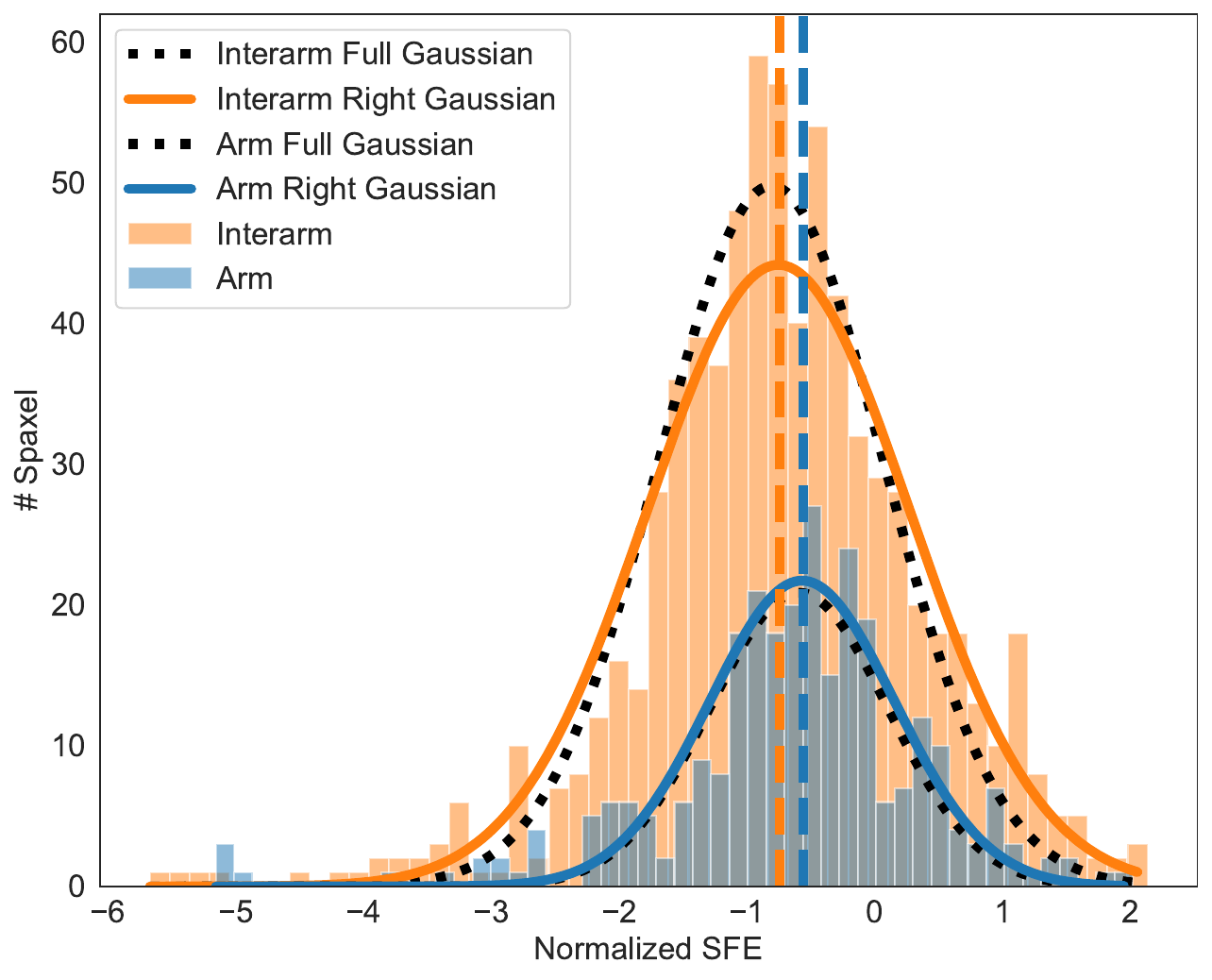}
    \caption{The molecular gas mass M$_\textup{gas}$ and star-formation efficiency (SFE) across six galaxies after correcting for inclinations. Left: The median of M$_\textup{gas}$ in arm and interarm regions are --0.124 and -0.301, respectively, the median  ratio is 1.505. Right: the SFE median in arm and interarm regions are  -0.560 and -0.745, corresponding to a median ratio of 1.533, consistent with the gatherer scenario.}
    \label{fig:Mgas_SFE_six_galaxies_combined}
\end{figure}

In Figures \ref{fig:sSFR_six_galaxies},  \ref{fig:SFR_Mstar_six_galaxies}, \ref{fig:Mgas_SFE_six_galaxies_combined}, we show the spaxel number distributions of all five  parameters. To compare the median differences with the width of each distribution, we fit a simple gaussian to the arm and interarm spaxel distributions, respectively, indicated by the dotted black lines in all plots. As a consistency check, we also fit the right side of each distribution only (since the left side may have a long tail), centered on the median values, to verify that our assumption of symmetric gaussians does not affect our results. The statistics are summarized in Table \ref{table:all_galaxies}. 
We find that, for all parameters, the differences between the median values are much smaller than the 1-$\sigma$ width of each distribution.

Our results indicate that there is no large difference between the median sSFR of the arm spaxels and that of interarm spaxels, consistent with the results of the individual galaxies. The six--galaxies median ratio between arm and interarms is still $<$2. The same holds for the median ratio of SFE.

\begin{table}[tbp]
\centering
\begin{tabular}{ l|c|c|c|c|c|c|c } 
\hline
Parameter & \# Arm & Median &  \# Interarm & Median & Median & Sigma  & Sigma\\
& spaxels & (log)  & spaxels & (log) & ratio & (Arm) & (Interarm)\\
\hline
Normalized SFR$_\textup{10Myr}$ & 374 & -0.636 & 1532 & -1.218 & 3.823 &0.642 & 1.007\\
Normalized $M_\textup{star}$ &374 & -0.061 & 1532 & -0.424 & 2.304 & 0.476 & 0.477\\
Normalized sSFR$_\textup{10Myr}$ & 374 & -0.682 & 1532 & -0.937 & 1.800 & 0.534 & 0.919\\
Normalized M$_\textup{gas}$ & 283 & -0.124 & 744 & -0.301 & 1.505 & 0.699 & 0.462 \\
Normalized SFE & 283 & -0.563 & 744 & -0.756 & 1.533 & 0.699 & 0.996\\
\hline
\end{tabular}
\caption{The medians, median ratios, and the width of gaussian curve fitting of arm/interarm spaxels. These are the spaxels of all galaxies, divided into two groups: arm and interarm spaxels. The gaussian curve is the single gaussian fit from median to the right.}
\label{table:all_galaxies}
\end{table}

\subsection{ Radial Trends}
\begin{figure}
    \centering
    \includegraphics[width=0.8\linewidth]{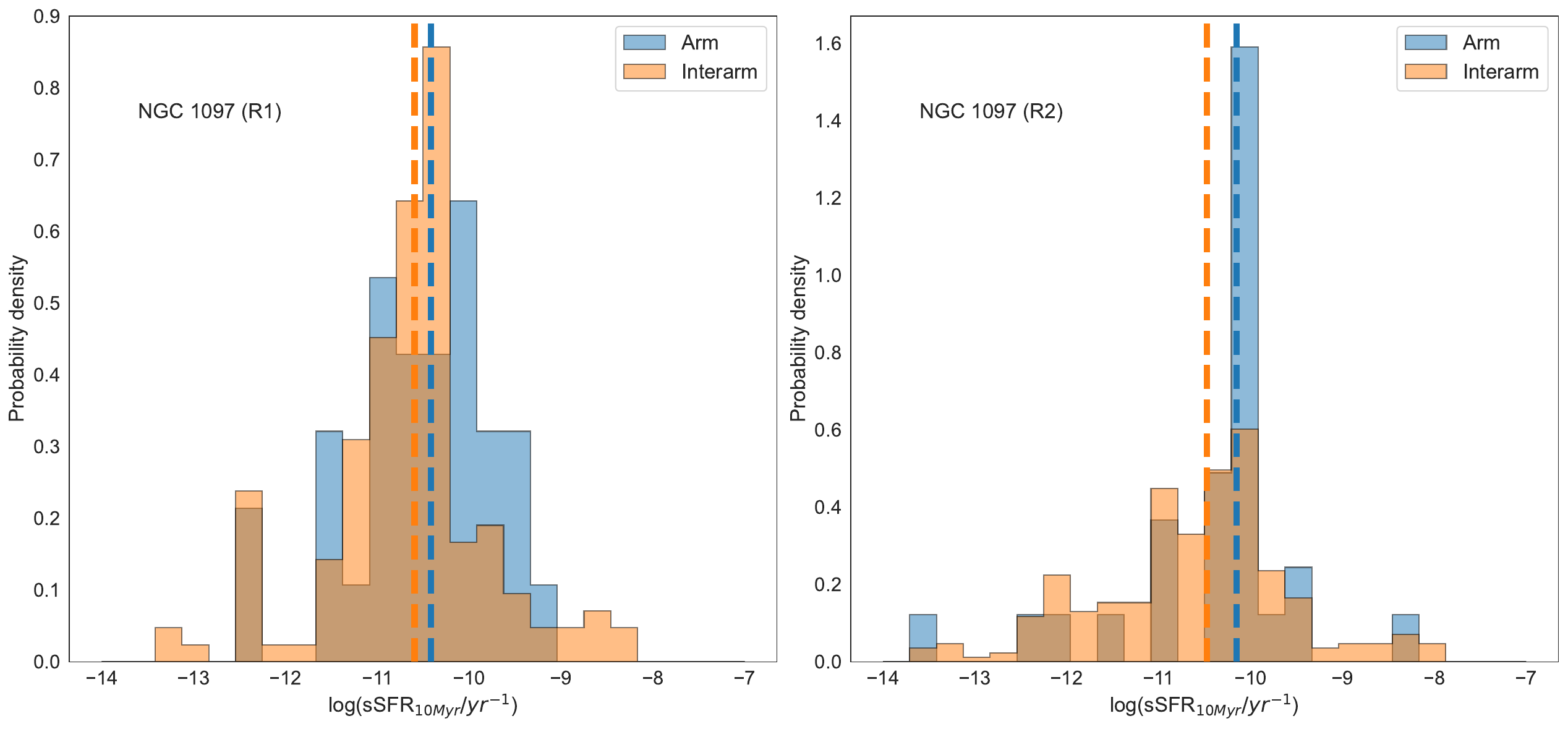}
    \includegraphics[width=0.8\linewidth]{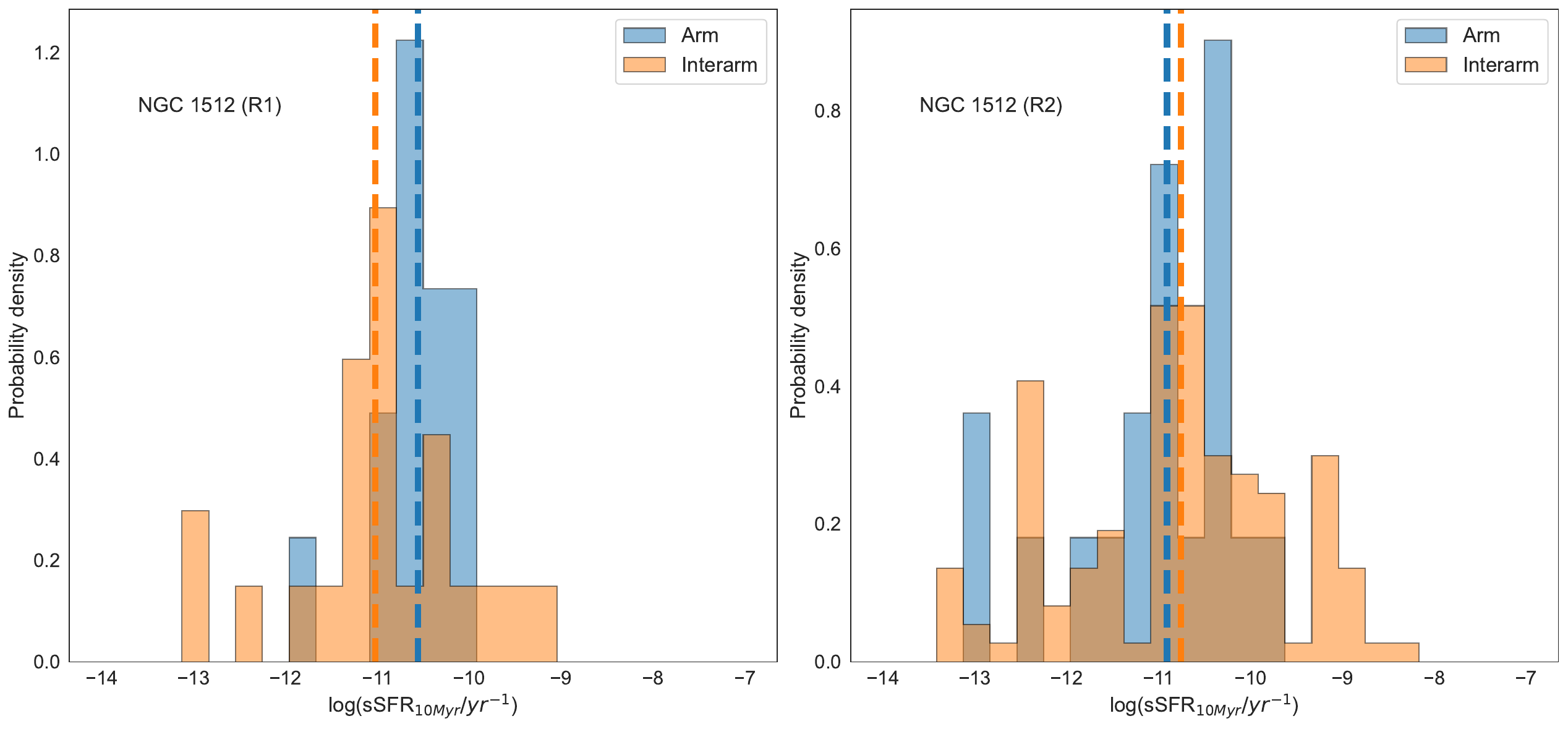}
    \includegraphics[width=0.8\linewidth]{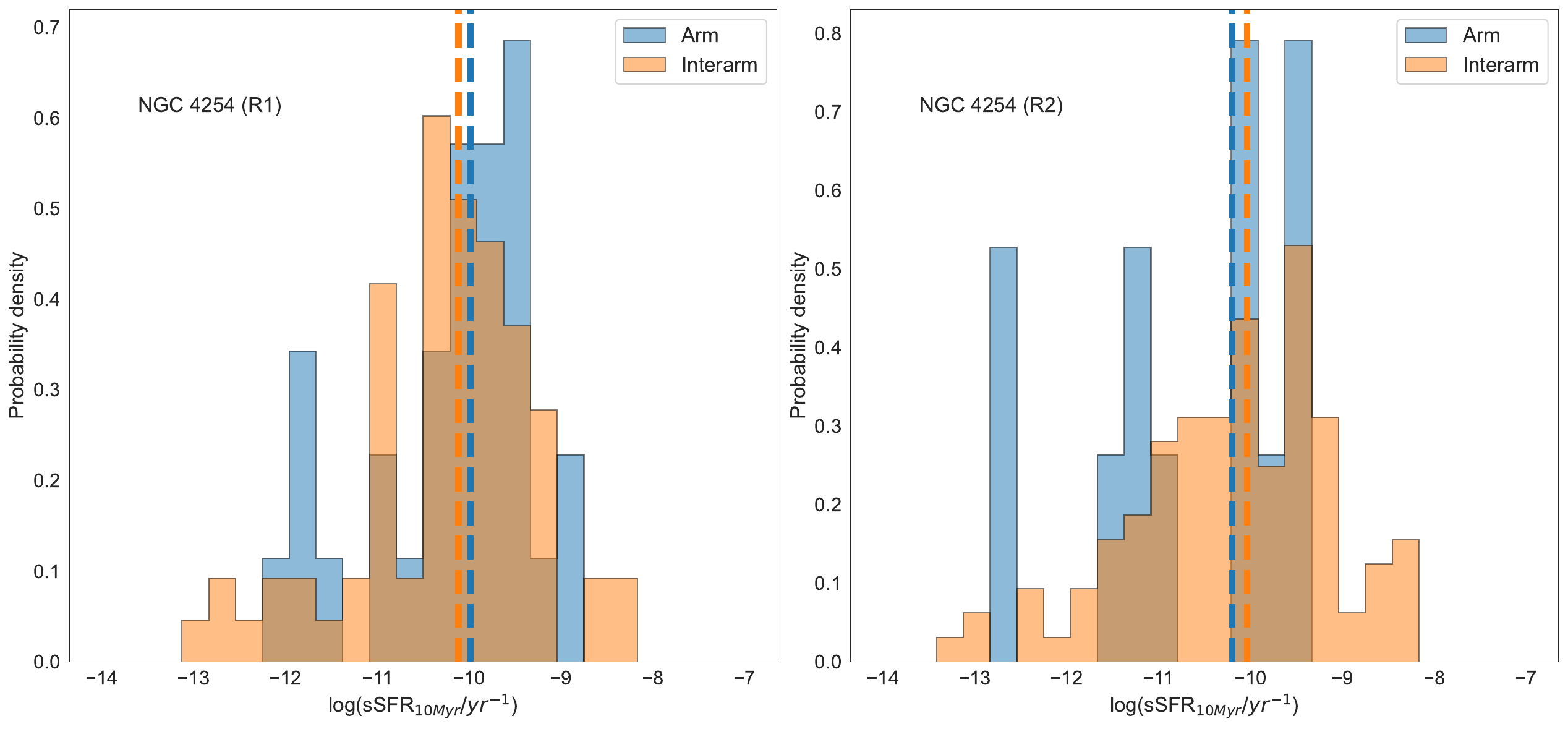}
    
    \caption{The sSFR distribution for arm and interarm spaxels in inner radius R1 (left column) and outer radius R2 (right column) for galaxy NGC 1097, NGC 1512, and NGC 4254. The median ratios are 1.507 and 2.104 for NGC 1097; 2.898 and 0.706 for NGC 1512; 1.357 and 0.691 for NGC 4254. The two small median ratios could be the result of small statistics, and their p-values are 0.4 and 0.8, meaning that these differences are not significant.}
    \label{fig:R1_R2_combined_sSFR-a}
\end{figure}

\begin{figure}
    \centering
    \includegraphics[width=0.8\linewidth]{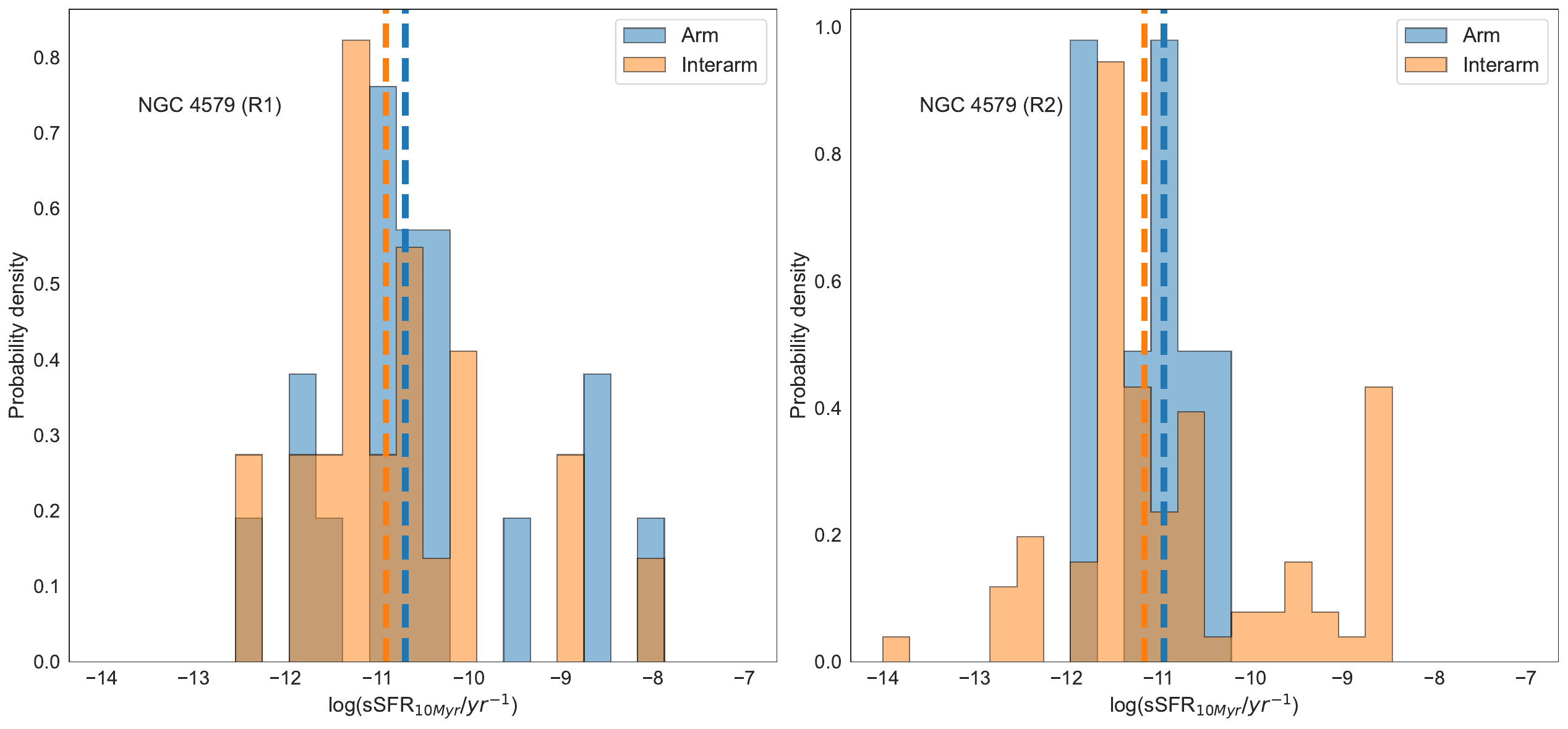}
    \includegraphics[width=0.8\linewidth]{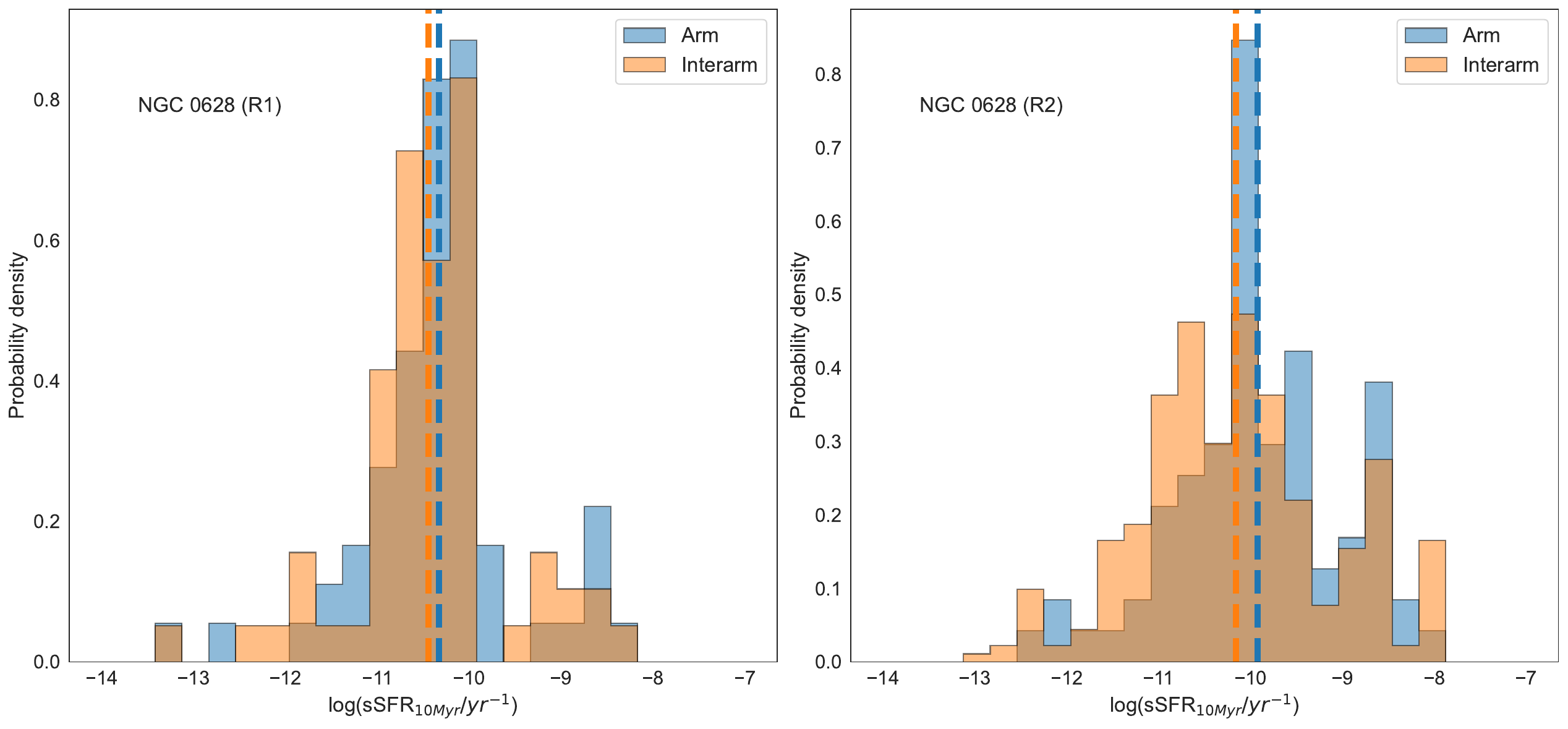}
    \includegraphics[width=0.8\linewidth]{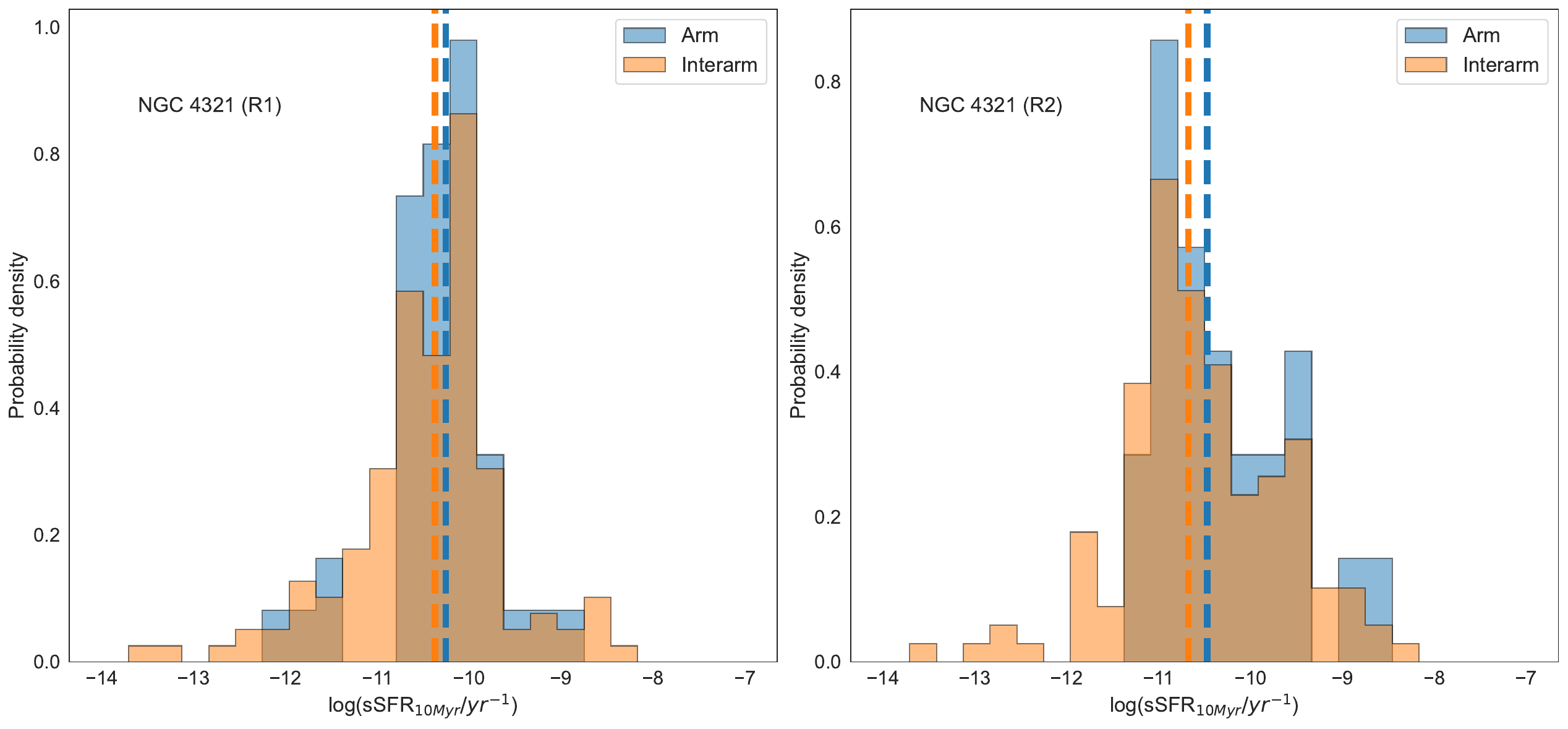}
    
    \caption{Similar to \ref{fig:R1_R2_combined_sSFR-a} but for galaxy NGC 4579, NGC 0628 and NGC 4321. The smallest median ratio is NGC 0628 (R1), 1.073; the largest median ratio is NGC 4579 (R1), 1.639. All median ratios are less than 2, and have p-values $>0.06$). }
    \label{fig:R1_R2_combined_sSFR-b}
\end{figure}

Similarly to what was presented in \citet{Sun+2024}, we test whether our results may depend on galactocentric distance, since several parameters (e.g., SFR and M$_\textup{star}$, Figure~\ref{fig:ngc1097-param-maps}) depend on that distance. We divide each galaxy into two radial bins, marked as R1 (inner) and R2 (outer). While NGC 628 and NGC 4321 are sufficiently extended that they could be divided into more than two radial bins as done in \citet{Sun+2024}, the other four galaxies contain too few spaxels to be divided into more than two radial bins. We determine the radial extent of each annulus by first identifying the outermost radius with a valid spaxel and then dividing this into two equal annuli. For NGC 4579 and NGC 1512, where the spiral arms are confined to the central region of the disk, we reduce the size of the inner annulus so that spiral arms in the outer annulus include about 10 spaxels or more. Even with this adjustment, the arm regions in R2 of NGC 4579 could still be affected by small number statistics. Figures~\ref{fig:R1_R2_combined_sSFR-a} and \ref{fig:R1_R2_combined_sSFR-b} show the distributions of the sSFR for the arm and interarm regions in each radial bin for each galaxy. The numerical values for all statistics are reported in Table~\ref{tab:Mstar-SFR-sSFR-2annuli} in Appendix~\ref{appendixB}. The radial analysis confirms our earlier statement that the differences in sSFR between arm and interarm regions are small, and non significant. We do not combine the radial bins into a super--galaxy, since the criterion we use to define each annulus is dictated by number statistics and not by the physical properties of the galaxies within each annular region.

\subsection{Comparison with Previous Work}
Our main result is that we do not find significant variations in the sSFR or the SFE between arm and interarm regions for the six individual galaxies, except for NGC 1097, where the sSFRs are different at the 3 $\sigma$ level, and NGC 628, which has a median SFE ratio $\sim$2 with a 3.9~$\sigma$ significance (Table \ref{table:individual_galaxies}). 
When we combine the parameters of all six galaxies together, the difference in sSFR medians between arm and interarm is only a factor of 1.8. 
There is no significant difference in the SFE either, with a median ratio of 1.5 between arm and interarm. Both results, for sSFR and SFE, are consistent with our earlier results on two galaxies \citep{Sun+2024}.
 
Our results can be compared with several previous studies, also leveraging the tight correlation between sSFR and SFE we find for our resolved galaxy regions (Figure \ref{fig:SFE_vs_sSFR_combined}).  
In particular, \citet{Elmegreen+1986,Foyle+2010,Kreckel+2016,Querejeta+2021,Querejeta+2024,Hart+2017} find little excess in the SFE in the arm regions compared to interarm regions, which is consistent with our findings.

For the SFR, we find contrasts around factors of 1.3–4 from the SED fits. This is in agreement with recipe-based results \citep{Foyle+2010, Querejeta+2024,Chen+2024}. We observe arm/interarm contrasts in stellar mass M$_\textup{star}$ being factors of 1.6-3.5, agreeing with \cite{Meidt+2021}.

As previously mentioned, simulation studies indicate that the presence of spiral arms enhances the sSFRs  by no more than a factor of 2, and this is considered evidence for the `gatherer' scenario, in which the primary role of spiral arms is to collect material rather than directly trigger star formation\citep{Dobbs+2011, Kim+2020}. In this light, our results support such scenario, with the only possible exception of NGC\,1097.

\subsection{Uncertainties and Limitations}
As mentioned in \cite{Sun+2024}, one major assumption in the SED-fitting method used in our study is that the stellar initial mass function (IMF) is constant and universal, from galaxy to galaxy, and between arm and interarm regions within galaxies. The other major assumption is that the energy balance approach can be applied to regions of scale $\sim 1$ kpc.

There is a continuing debate about whether the IMF is constant and universal \citep{Bastian+2010, Jung+2023}. Studies have shown that under the assumption of a universal Chabrier IMF, non-constant star formation histories (SFHs) are capable of explaining the observed flux distributions of galaxies, especially in low-mass ones \citep{Weisz+2012}. Star clusters in both high-- and low--mass galaxies are consistent with assumptions of a universal IMF as well \citep[e.g.,][ and references therein]{Jung+2023}. 
We also ensure that there is no small-number statistics problem in most regions, where $M_\textup{star}\gg 10^6 M_\odot$. Thus, we consider the assumption of a universal IMF to be reasonable when studying individual galaxy regions.

Previous work \citep{Smith&Hayward2018, Battisti+2019} tests the energy balance assumption and concludes that it is reasonable when considering scales that are $\gtrsim1$~kpc. Ionizing photons escaping from HII regions are known to travel distances ranging from a few hundred parsecs to around one kiloparsec within galaxy disks \citep{Haffner+2009}. Similarly, infrared (IR) emission from dust is typically confined to within several hundred parsecs of the corresponding heating source \citep{Lawton+2010}. This allows us to assume energy balance in each region within a galaxy.

\section{Summary and Conclusions}\label{sec:summary}
In this work, we extend our previous study of two galaxies to six diverse local galaxies, and investigate their SFRs, M$_\textup{star}$, M$_\textup{gas}$, sSFRs, and SFE. We divide each galaxy into a few tens to hundreds of regions with spatial scale $\sim 1-1.5$ kpc. We construct the SED for individual regions, using about 19–20 broadband photometric data covering wavelength range from FUV to FIR. We use the MAGPHYS fitting code for SED fitting, deriving the SFRs, M$_\textup{star}$ and, from these, sSFRs. We also use publicly available CO maps of the galaxies to derive M$_\textup{gas}$ and SFE.

We first extend the study of \citet{Saintonge+2011}, where they find a correlation between sSFR and SFE for local galaxies, to the $\sim$kpc--size regions in our six galaxies. We determine that sSFR and SFE are tightly correlated also for sub--galactic regions. Thus, sSFR and SFE can be used interchangeably to investigate properties of spiral arms, which we plan to use in future studies.

We find that for all six galaxies, arm regions have higher stellar mass, molecular gas mass, and SFR than those in the interarm regions. However, the medians and distributions of sSFR and SFE are almost indistinguishable between arm and interarm regions, with median ratios $\lesssim 2$ and p-values $\geq 0.05$ for five of our galaxies. NGC1097 and NGC 628 provide limited exceptions, with contrasts between arm and interarm regions of a factor of $\sim$2 in sSFR and SFE, respectively, but at a moderate level of significance ($\sim$3~$\sigma$ and 3.9~$\sigma$, respectively). 

We also combine the best-fit parameters of the six galaxies together. The median ratio of arm and interarm sSFR and SFE are $\sim$1.8 and $\sim$1.5, respectively, much smaller than the width of the parameter distributions for the two types of regions. 
Thus we conclude that the result for the combined sample is consistent with the results of individual galaxies, which supports the `gatherers' scenario for the role of spiral arms in galaxies. However, the moderate deviations we observe in NGC 1097 and NGC 628 encourage further investigations, which we plan on pursuing on a larger sample to understand whether some galaxies may have properties that differentiate them from others.

\appendix

\section{Distributions of parameters in individual galaxies}\label{appendixA}

\begin{figure}
    \centering
    \includegraphics[width=1\linewidth]{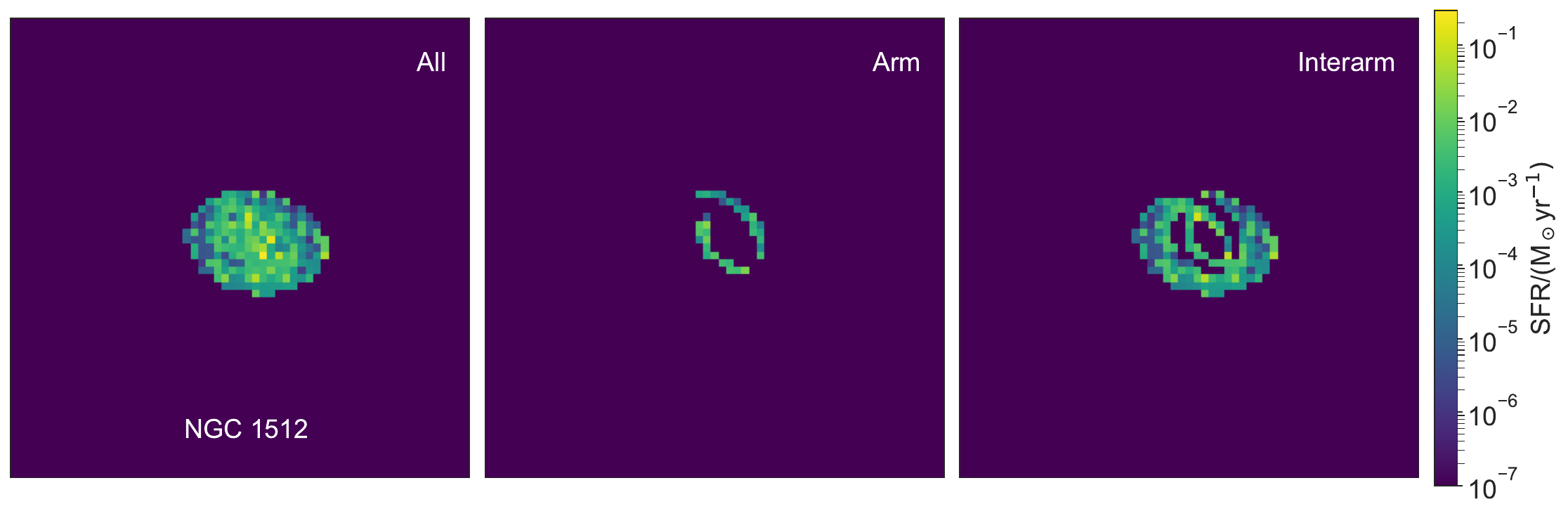}
    \includegraphics[width=1\linewidth]{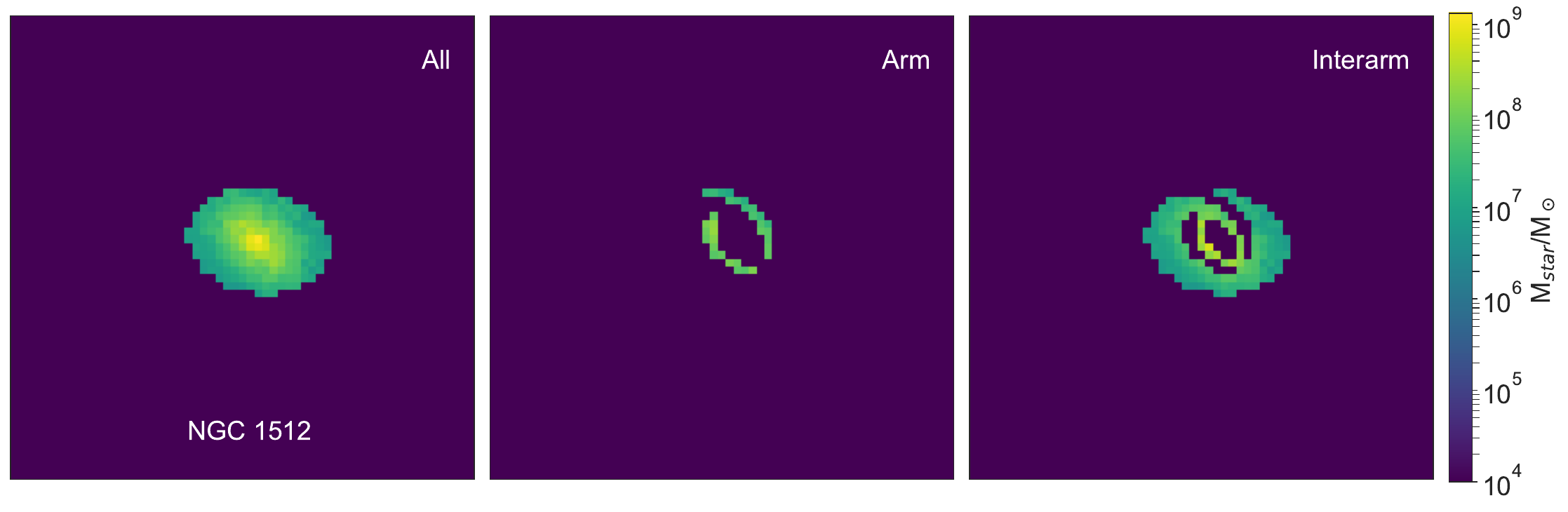}
     \includegraphics[width=1\linewidth]{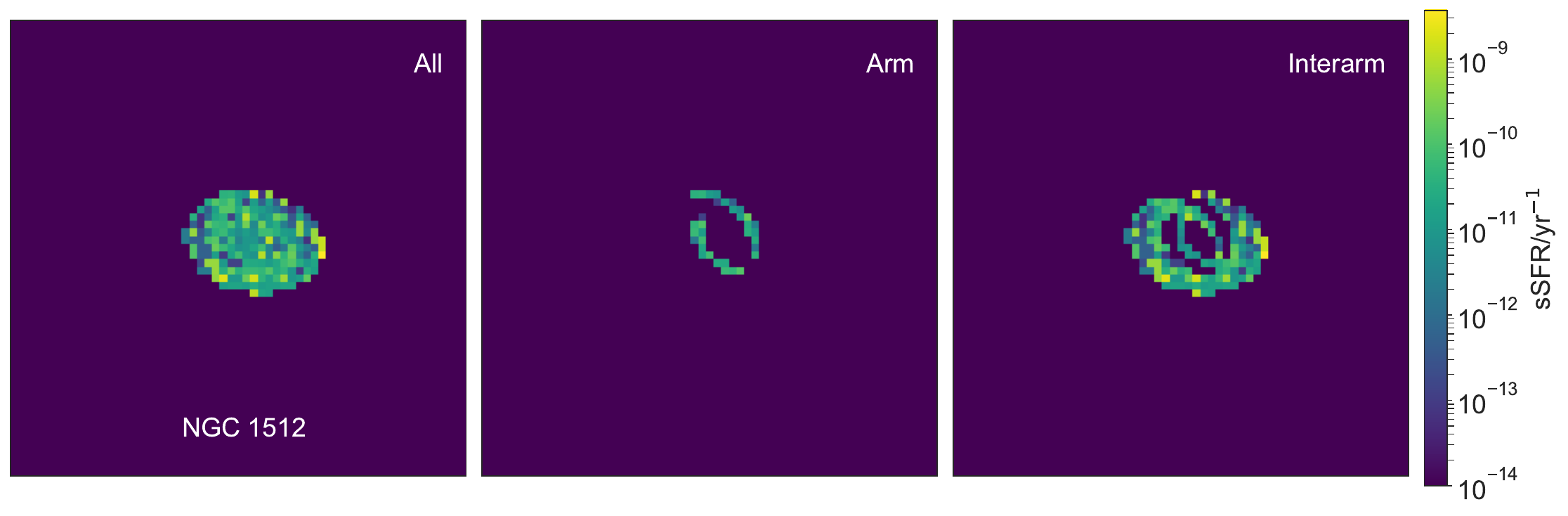}
    \caption{Similar to Figure \ref{fig:ngc1097-param-maps} but for NGC 1512. 
    }
    \label{fig:ngc1512-param-maps}
\end{figure}

\begin{figure}
    \centering
    \includegraphics[width=1\linewidth]{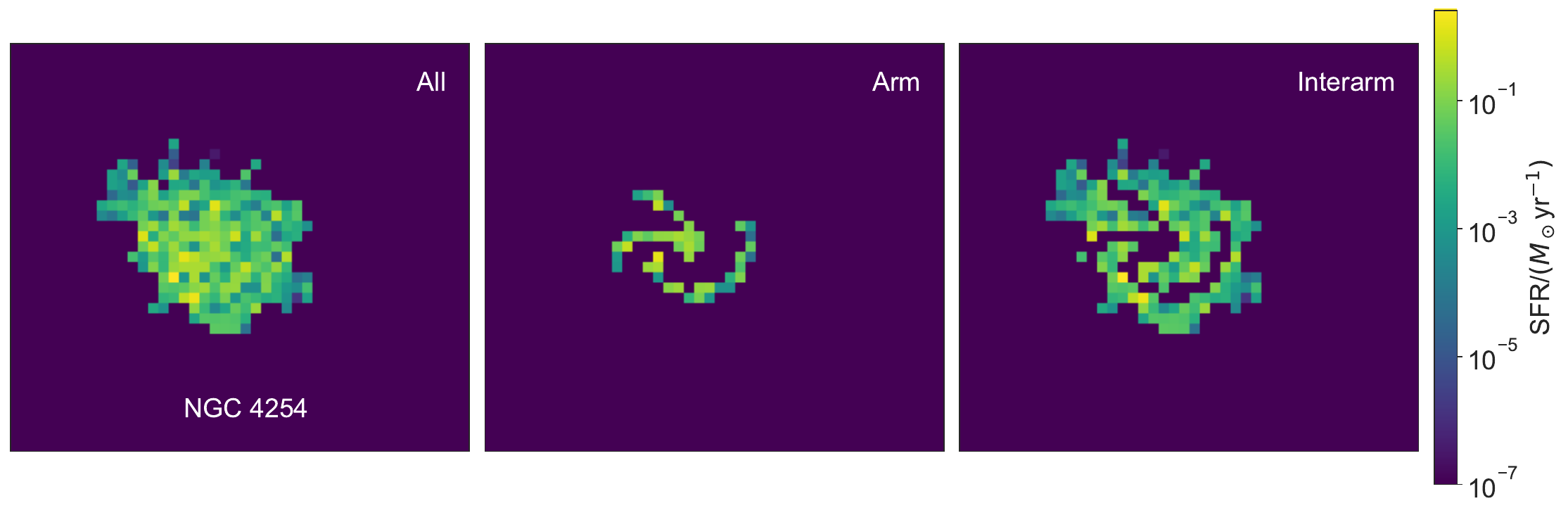}
    \includegraphics[width=1\linewidth]{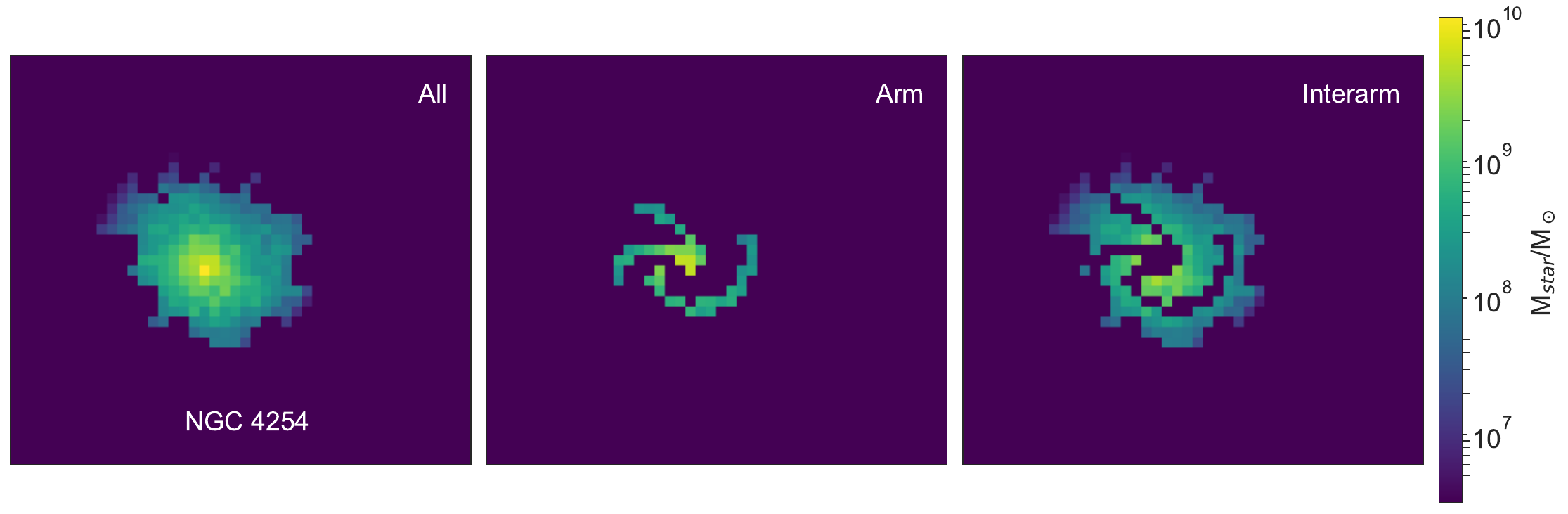}
     \includegraphics[width=1\linewidth]{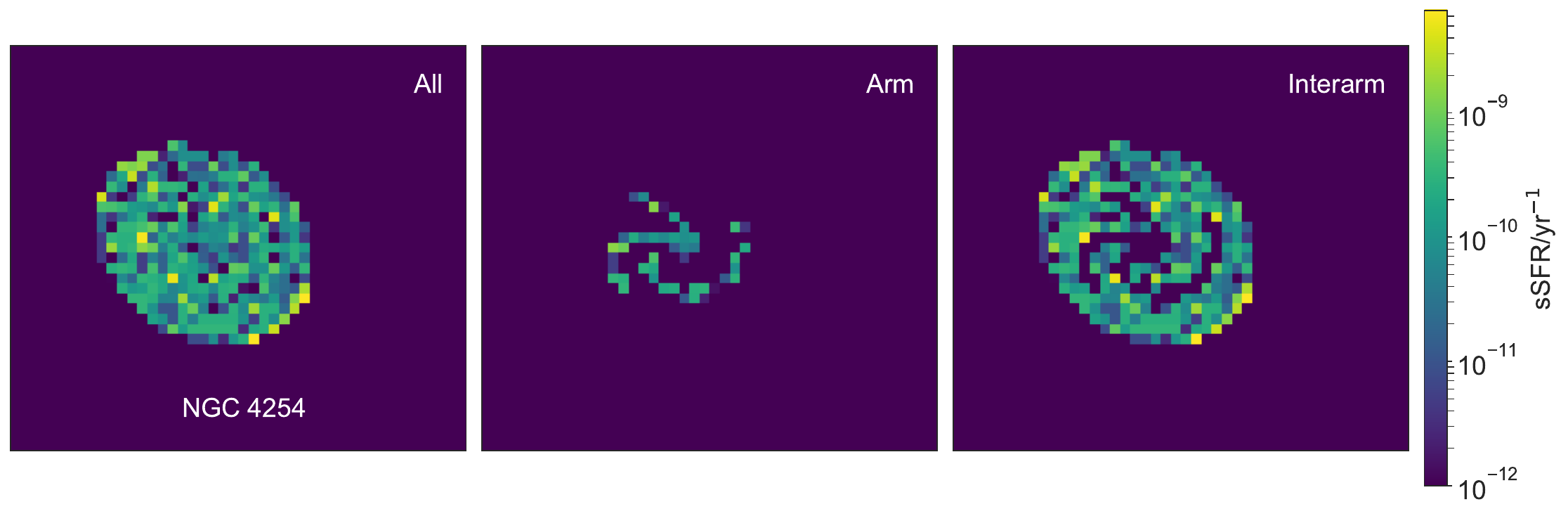}
    \caption{Similar to Figure \ref{fig:ngc1097-param-maps} but for NGC 4254. In each row, the three panels from left to right are: best-fit values of all spaxels with S/N 3 in SPIRE 250 image, including bulge spaxels; arm spaxels; interarm spaxels, with the bulge excluded.}
    \label{fig:ngc4254-param-maps}
\end{figure}

\begin{figure}
    \centering
    \includegraphics[width=1\linewidth]{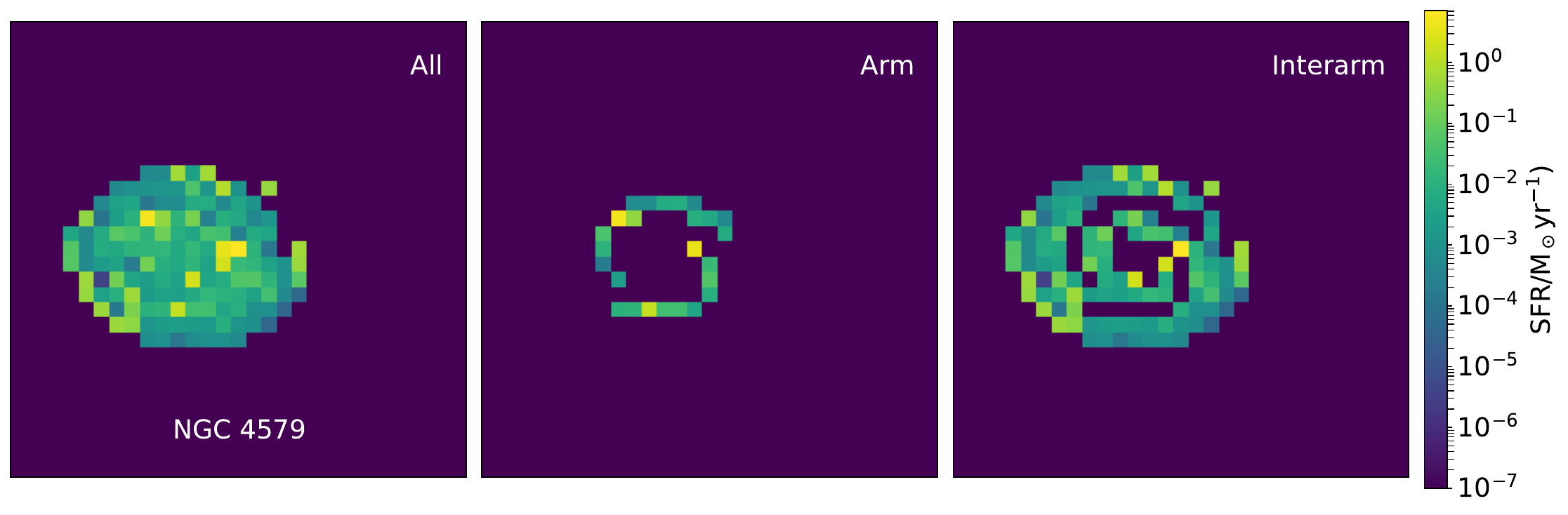}
    \includegraphics[width=1\linewidth]{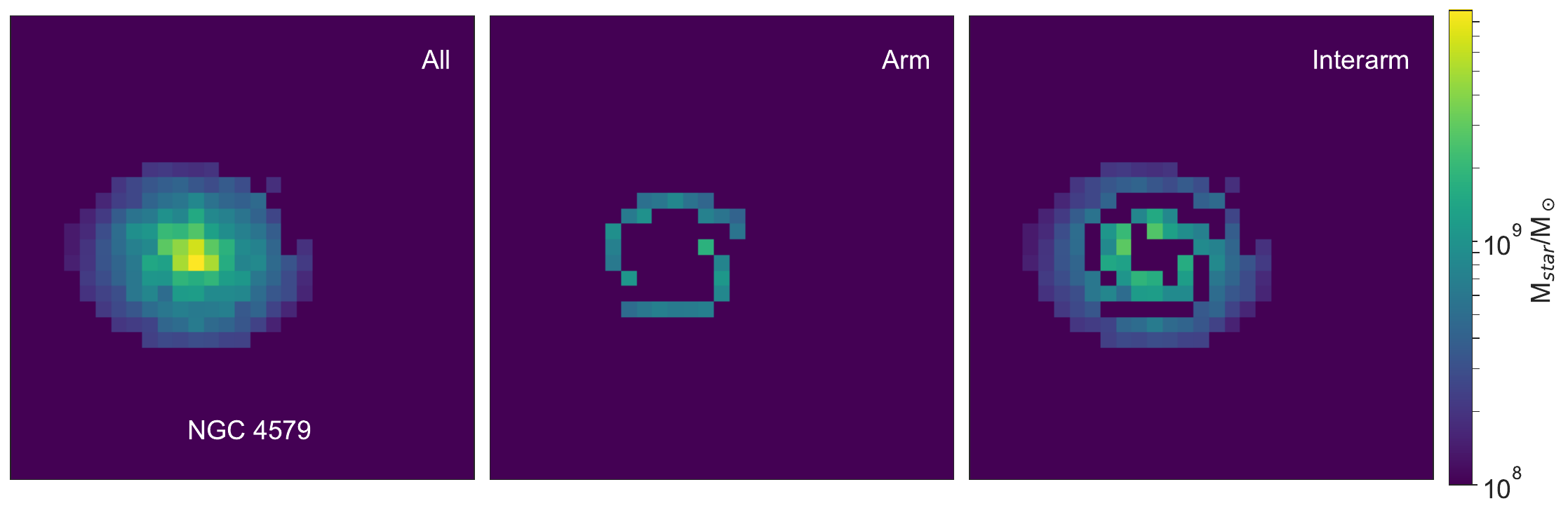}
     \includegraphics[width=1\linewidth]{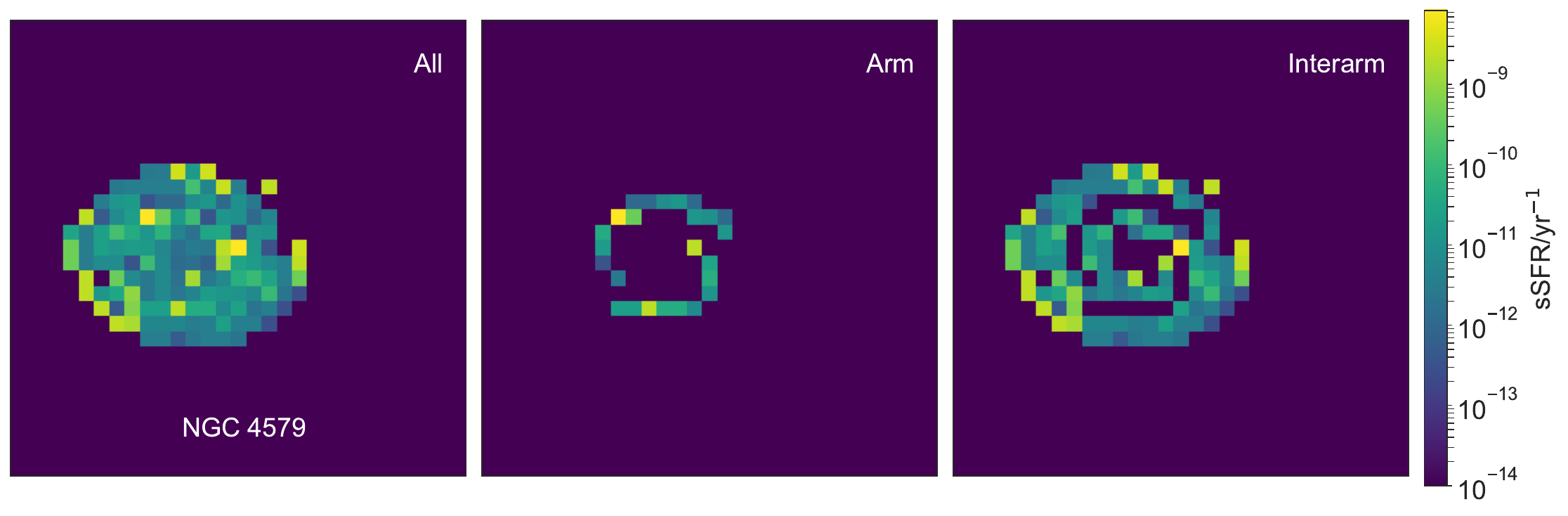}
    \caption{Similar to Figure \ref{fig:ngc4254-param-maps} but for NGC 4579. In each row, the three panels from left to right are: best-fit values of all spaxels with S/N 3 in SPIRE 250 image, including bulge spaxels; arm spaxels; interarm spaxels, with the bulge excluded.
    }
    \label{fig:ngc4579-param-maps}
\end{figure}

\begin{figure}
    \centering
    \includegraphics[width=1\linewidth]{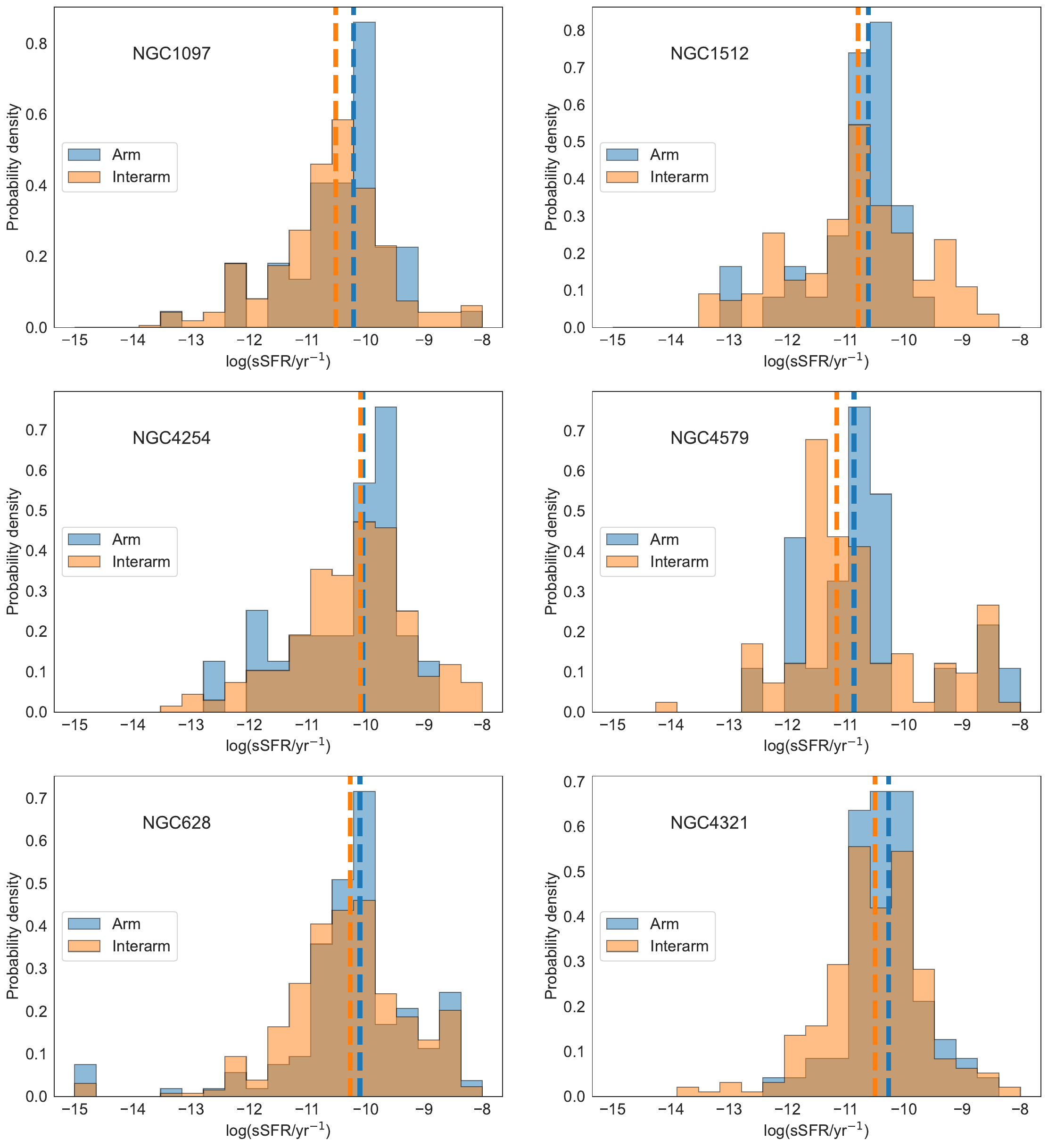}
    \caption{Distributions of best-fit values of the sSFR (where the SFR is calculated over the most recent $10^7$ yr) for the six galaxies. The arm spaxels and the interarm spaxels are plotted in blue and orange, respectively. For both galaxies, the sSFR distributions of arm/interarm regions are similar and their median values are shown as vertical dashed lines.}
    \label{fig:six-sSFR-hists}
\end{figure}

\begin{figure}
    \centering
     \includegraphics[width=1\linewidth]{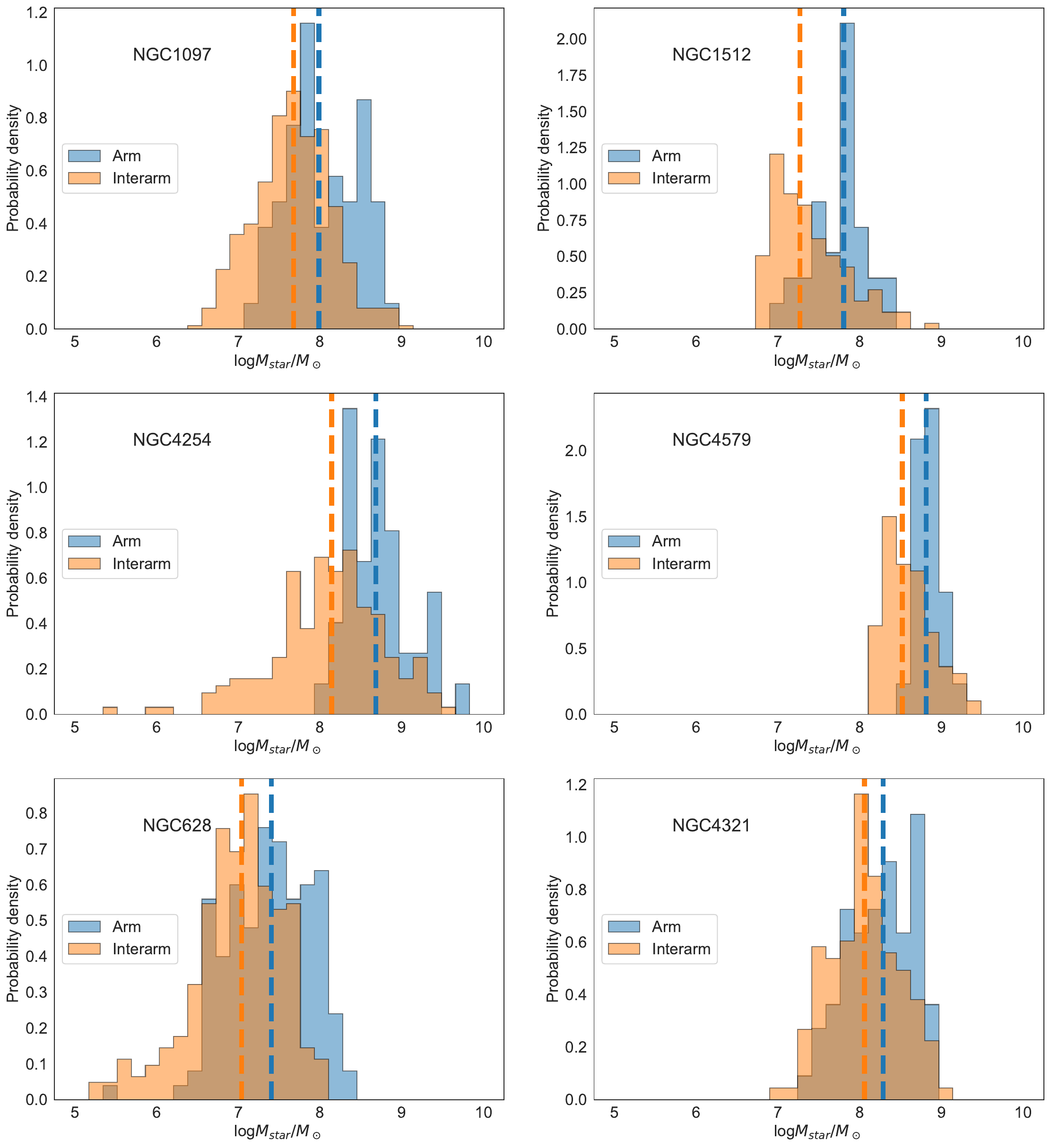}
    \caption{Similar to Figure \ref{fig:six-sSFR-hists} but for M$_\textup{star}$.}
    \label{fig:six-Mstar-hists}
\end{figure}

\begin{figure}
    \centering
     \includegraphics[width=1\linewidth]{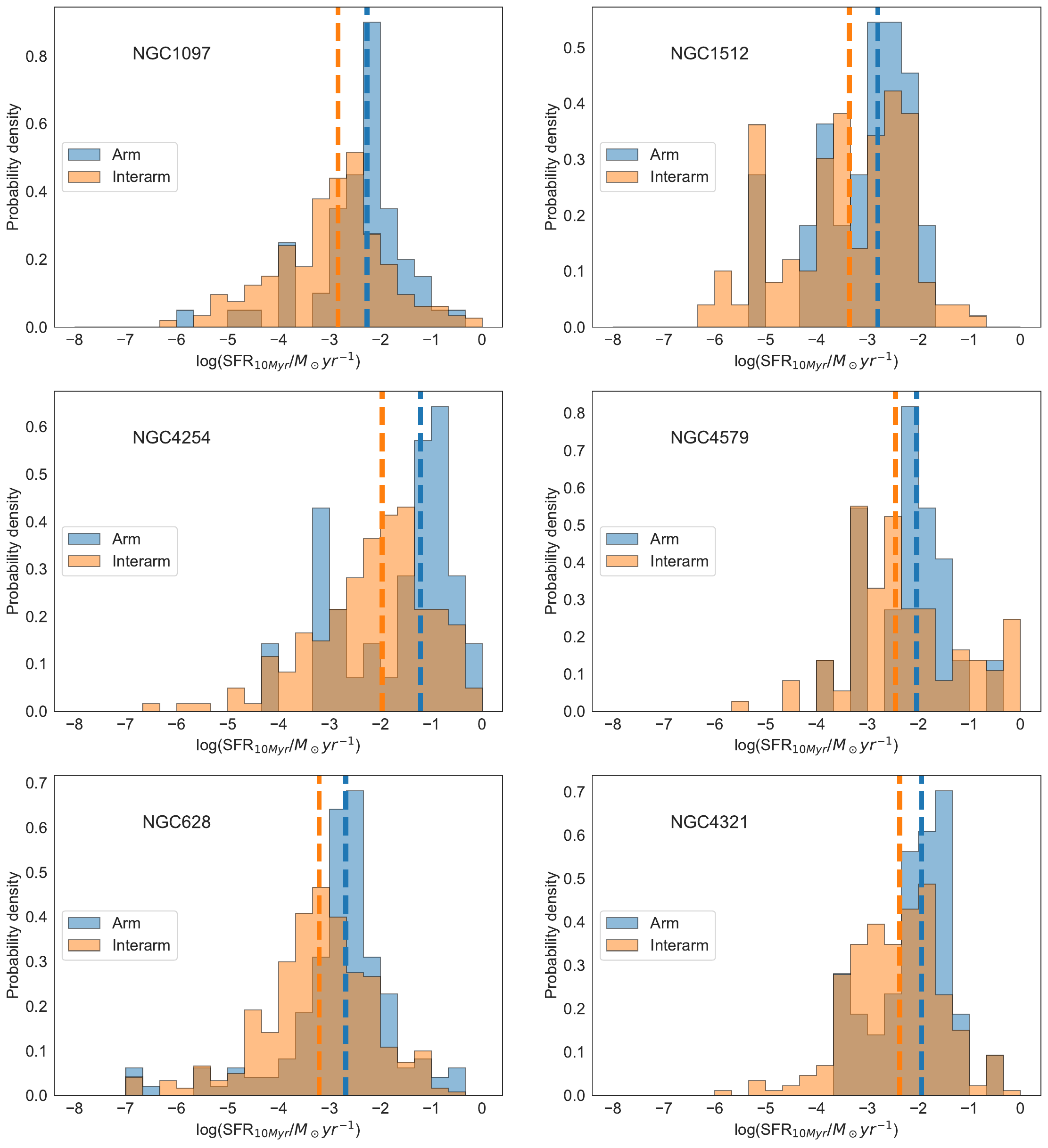}
    \caption{Similar to Figure \ref{fig:six-sSFR-hists}, but for SFR$_\textup{10Myr}$.}
    \label{fig:six-SFR-hists}
\end{figure}

The physical parameter maps (SFR, M$_\textup{star}$ and sSFR) for NGC 1512, NGC 4254 and NGC 4579 are shown in Figures \ref{fig:ngc1512-param-maps}, \ref{fig:ngc4254-param-maps}, and \ref{fig:ngc4579-param-maps}, while the distributions of sSFR spaxels for the six individual galaxies are given in Figure  \ref{fig:six-sSFR-hists}. In all panels of this Figure, the arm spaxels (in blue) are separated from the interarm spaxels (in orange), and the median of each distribution is shown as a vertical dashed line. 
The median ratios between arm/interarm spaxels are $\lesssim 2$ for all galaxies. 
See table \ref{table:individual_galaxies} for the exact values of medians and the KS-test results of each distribution. Similarly, the six distributions of M$_\textup{star}$ and SFR are shown in Figures \ref{fig:six-Mstar-hists} and \ref{fig:six-SFR-hists}, respectively. 

\section{Statistics in Radial Bins}\label{appendixB}
We divide each galaxy into two radial bins, marked as R1 (inner) and R2 (outer). Table \ref{tab:Mstar-SFR-sSFR-2annuli} shows the statistics of arm and interarm regions in each radial bin for each of the six galaxies.

\begin{table}[]
    \centering
    \begin{tabular}{c|c|c|c|c|c|c|c}
    \hline
    Annulus & Parameter & Arm & Interarm &  Arm & Interarm & Median ratio &  KS-test\\
 & & \#spaxels  & \#spaxels & median (log) & median (log) & (Arm/Interarm) & p-value\\
    \hline
    NGC 1097\\
    \hline
    R1 & M$_\textup{star}$ & 32 & 144 & 8.395 & 8.089 & 2.020 & 3.571e-4\\
    R2 & M$_\textup{star}$ & 28 & 290& 7.675 & 7.498 & 1.503 & 0.006\\
    R1 & SFR$_\textup{10Myr}$ &  32 & 144& -2.070 & -2.462 & 2.466 & 0.071\\
    R2 & SFR$_\textup{10Myr}$ & 28 & 290& -2.340 & -3.050 & 5.133 & 3.560e-4\\
    R1 & sSFR & 32 & 144 & -10.416 & -10.594 & 1.507 & 0.078\\
    R2 & sSFR & 28 & 290 & -10.150 & -10.473 & 2.104 & 0.005\\
    \hline
    NGC 1512\\
    \hline
    R1 & $M_\textup{star}$ & 14 & 23 & 7.885 & 8.112 & 0.593 & 0.038\\
    R2 & $M_\textup{star}$ & 19 & 126 & 7.552 & 7.207 & 2.210 & 1.373e-4\\
    R1 & SFR$_\textup{10Myr}$ & 14 & 23 & -2.590 & -2.654 & 1.159 & 0.699\\
    R2 & SFR$_\textup{10Myr}$ & 19 & 126 & -3.414 & -3.455 & 1.098 & 0.866\\
    R1 & sSFR & 14 & 23 & -10.558 & -11.020 & 2.898 & 0.0317\\
    R2 & sSFR & 19 & 126 & -10.909 & -10.758 & 0.706 & 0.434\\
    \hline
    NGC 4254 \\
    \hline
        R1 & $M_\textup{star}$ & 30 & 74 & 8.808 & 8.636 & 1.487 & 0.023\\
        R2 & $M_\textup{star}$ & 13 & 110 & 8.327 & 7.784 & 3.490 & 3.821e-08\\
        R1 & SFR$_\textup{10Myr}$ & 30 & 74 & -0.960 & -1.523 & 3.646 & 0.012\\
        R2 & SFR$_\textup{10Myr}$ &  13 & 110 & -1.931 & -2.341 & 2.575 & 0.355\\
        R1 & sSFR & 30 & 74 & -9.984 & -10.117 & 1.357 & 0.813\\
        R2 & sSFR & 13 & 110 & -10.197 & -10.036 & 0.691 & 0.800\\
    \hline
    NGC 4579\\
    \hline
    R1 & $M_\textup{star}$ &18 & 25 & 8.827 & 9.067 & 0.576 & 1.275e-4\\
    R2 & $M_\textup{star}$ & 7 & 87 & 8.753 & 8.452 & 2.000 & 4.979e-4\\
    R1 & SFR$_\textup{10Myr}$ & 18 & 25 & -1.882 & -1.882 & 0.999 & 0.986\\
    R2 & SFR$_\textup{10Myr}$ & 7 & 87 & -2.156 & -2.586 & 2.696 & 0.336\\
    R1 & sSFR & 18 & 25 & -10.694 & -10.908 & 1.639 & 0.409\\
    R2 & sSFR & 7 & 87 & -10.942 & -11.154 & 1.629 & 0.714\\
   \hline
   NGC 0628\\
   \hline
   R1 & $M_\textup{star}$ & 66 & 69 & 7.824 & 7.665 & 1.442 & 1.080e-4\\
   R2 & $M_\textup{star}$ & 81 & 312 & 6.996 & 6.883 & 1.296 & 0.008\\
   R1 & SFR$_\textup{10Myr}$ & 66 & 69 & -2.567 & -2.791 & 1.675 & 0.004\\
   R2 & SFR$_\textup{10Myr}$ & 81 & 312 & -2.846 & -3.321 & 2.983 & 2.517e-6\\
   R1 & sSFR & 66 & 69 & -10.456 & -10.487 & 1.073 & 0.059\\
   R2 & sSFR & 81 & 312 & -10.369& -10.535 & 1.465 & 0.065\\
   \hline
   NGC 4321\\
   \hline
   R1 & $M_\textup{star}$ & 42 & 135 & 8.495 & 8.324 & 1.483& 0.031\\
   R2 & $M_\textup{star}$ & 24 & 134 & 7.837 & 7.709 & 1.342 & 0.404\\
   R1 & SFR$_\textup{10Myr}$ & 42 & 135 & -1.781 & -2.009 & 1.691 & 0.016\\
   R2 & SFR$_\textup{10Myr}$ & 24 & 134 & -2.589 & -2.910 & 2.094 & 0.366\\
   R1 & sSFR & 42 & 135 & -10.255 & -10.372 & 1.309 & 0.188\\
   R2 & sSFR & 24 & 134 & -10.470 & -10.677 & 1.611 & 0.374\\
   \hline
    \end{tabular}

   \caption{Medians, median ratios, and KS-tests of parameters in two annuli for the six galaxies. }
    \label{tab:Mstar-SFR-sSFR-2annuli}
\end{table}

\section{SFE vs sSFR in individual galaxies}\label{appendixC}
\begin{figure}
    \centering
    \includegraphics[width=0.9\linewidth]{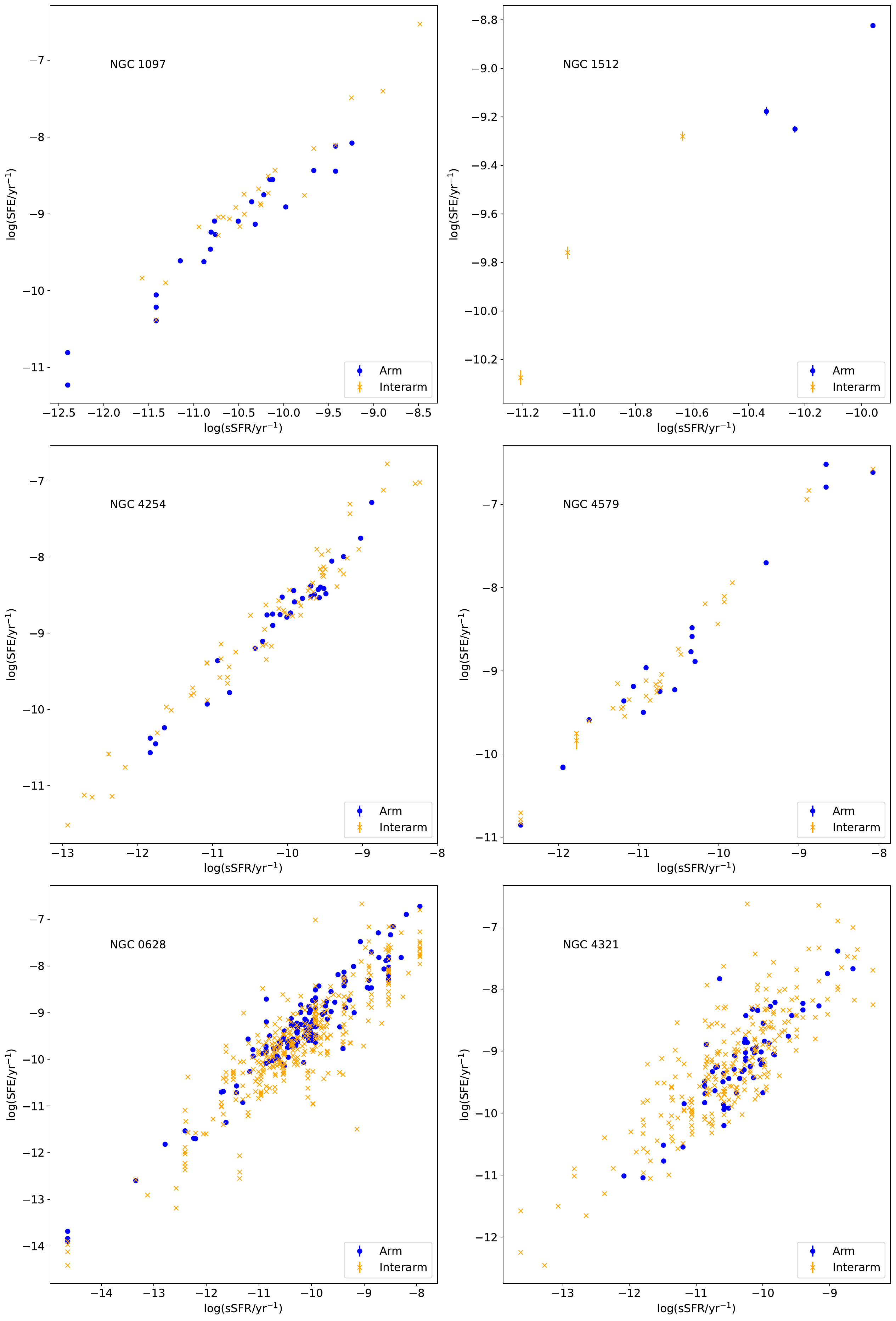}
    \caption{SFE vs sSFR for each galaxy.}
    \label{fig:SFE_vs_sSFR_subplots}
\end{figure}

We test if the relation of SFE and sSFR in \cite{Saintonge+2011} also holds for resolved, kpc--sized galaxy regions. We find tight correlations in individual galaxies as well as all galaxies when combined together (see main text). We show the results for individual galaxies here in Figure \ref{fig:SFE_vs_sSFR_subplots}. 

\section*{Acknowledgement}
The authors thank the anonymous referee for the many useful comments that have helped greatly improve this manuscript.

B.S. and D.C. acknowledge support from the grant NASA ADAP/2021, ID 80NSSC22K0478, `1,000 star  formation histories in nearby galaxies' for this research.

This research has made use of the NASA/IPAC Extragalactic Database (NED) which is operated by the Jet Propulsion Laboratory, California Institute of Technology, under contract with the National Aeronautics and Space Administration.

\software{
astropy \citep{2013A&A...558A..33A},
numpy \citep{harris2020},
matplotlib \citep{matplotlib},
scipy \citep{2020SciPy-NMeth},
photutils \citep{photutils},
SWarp \citep{Bertin1996},  
MAGPHYS \citep{daCunha2008MNRAS}
Seaborn \citep{Waskom2021}
}

\bibliography{reference}{}
\bibliographystyle{aasjournal}

\end{document}